\documentclass[prd,twocolumn,preprintnumbers,superscriptaddress,amsmath,amssymb]{revtex4-1}
\usepackage{graphicx}
\usepackage{xcolor}

\begin{document}


\title{ Classification of the quantum chaos in colored Sachdev-Ye-Kitaev models  }
\author{Fadi Sun}
\affiliation{
Department of Physics and Astronomy,
Mississippi State University, MS 39762, USA}
\affiliation{University of Science and Technology Beijing, Beijing 100083, China }
\affiliation{
Kavli Institute of Theoretical Physics, University of California,
Santa Barbara, Santa Barbara, CA 93106, USA}

\author{Yu Yi-Xiang}
\affiliation{
Beijing National Laboratory for Condensed Matter Physics,
Institute of Physics, Chinese Academy of Sciences, Beijing 100190, China}

\author{Jinwu Ye}
\affiliation{
Department of Physics and Astronomy,
Mississippi State University, MS 39762, USA}
\affiliation{University of Science and Technology Beijing, Beijing 100083, China
}
\affiliation{
Kavli Institute of Theoretical Physics, University of California,
Santa Barbara, Santa Barbara, CA 93106, USA}

\author{W.M. Liu}
\affiliation{
Beijing National Laboratory for Condensed Matter Physics,
Institute of Physics, Chinese Academy of Sciences, Beijing 100190, China}

\begin{abstract}
The random matrix theory (RMT) can be used to classify
both topological phases of matter and quantum chaos.
We develop a systematic and transformative RMT
to classify the quantum chaos in the colored Sachdev-Ye-Kitaev (SYK) model
first introduced by Gross and Rosenhaus.
Here we focus  on the 2-colored case and 4-colored case with balanced number of Majorana fermion $ N $.
By identifying the maximal symmetries, the independent parity conservation sectors,
the minimum (irreducible) Hilbert space,
and especially the relevant anti-unitary and unitary operators,
we show that the color degree of freedoms lead to novel quantum chaotic behaviours.
When $ N $ is odd, different symmetry operators need
to be constructed to make the classifications complete.
The 2-colored case only show 3-fold Wigner-Dyson way,
and the 4-colored case show 10-fold generalized Wigner-Dyson way
which may also have non-trivial edge exponents.
We also study 2- and 4-colored hybrid SYK models
which display many salient quantum chaotic features hidden in the corresponding pure SYK models.
These features motivate us to develop a systematic RMT to study the energy level statistics of 2 or 4 un-correlated random matrix ensembles.
The exact diagonalizations are performed to study
both the bulk energy level statistics and the edge exponents
and find excellent agreements with our exact maximal symmetry classifications.
Our complete and systematic methods can be easily extended to study the generic imbalanced cases.
They may be transferred to the classifications of colored tensor models,
quantum chromodynamics  with pairings across different colors, quantum black holes
and interacting symmetry protected (or enriched) topological phases.
\end{abstract}

\maketitle

\section{Introduction}

Classifications of phases of matter has a long history.
It starts with the Landau theory on the classifications of all the possible
spontaneous symmetry breaking states to more recent classifications of
the topological insulators and superconductors
\cite{kane,zhang,tenfold} of non-interacting electrons
using the same symbols and techniques as the random matrix theory (RMT).
The latter inspired and triggered the classifications of topological phase
of interacting bosons or fermions which break no symmetries \cite{tenfold,wen}.
On the other forefront, there are recent extensive research activities
on studying quantum chaos and quantum information scramblings
in Sachdev-Ye-Kitaev (SYK) model \cite{SY,SY3,Kit,subir3,me} and its various invariants.
Because the ground state of SYK models are quantum spin liquids
which break neither symmetry nor having any kinds of topological orders.
Then one may need to classify the SYK models by a different organization pattern
of matter which describe how quantum information are scrambled in the system: the quantum chaos.

  There are two completely independent ways to characterize the quantum chaos.
  One way is to evaluate the out of time ordered correlation (OTOC) functions
  to describe the quantum information scramblings in the early time (Ehrenfest time)
  \cite{Pol,Mald,Gross,longtime,liu1,liu2,Kitnon,OTOCrev}.
  It was found that the SYK models show the maximal quantum  chaos with the largest possible Lyapunov exponent
  $ \lambda_L= 2 \pi/\beta $  saturating the quantum chaos bound \cite{chaosbound}.
  This salient feature ties that of the quantum black holes which are the fast quantum information scramblers in Nature.
  This fact suggests that the SYK model may be a boundary theory of some sort of bulk dilaton gravity theory such as the well known Jackiw-Teitelboim (JT) gravity \cite{Ads2Mald,Kitblack}.
  Another way is to use the RMT to describe the
  energy level statistics (ELS) which  can be used to probe the late time (Heisenberg time) dynamics
  \cite{CSYKnum,MBLSPT,randomM,randomM0,randomsusy1,randomsusy2,KAMSYK}.
  It was found that the ELS of the Majorana fermion SYK  can be described by the
  3 fold-way Wigner-Dyson distributions  in a $N \pmod 8$ periodicity \cite{MBLSPT,randomM0,randomM}.
  The RMT has also been employed to study the quantum chaotic behaviours of event horizon
  fluctuations of black holes \cite{randomM}.

The quantum chaos in the SYK models are due to the quenched disorders.
However, it inspired a new class of clean quantum mechanical models
called colored (Gurau-Witten) or un-colored (Klebanov-Tarnopolsky) tensor model
\cite{tensor1,tensor2,tensor3,graph,tensorSYK1,tensorrev}
which share similar quantum chaotic properties as the SYK at least in the large $N$ limit
\cite{uncolor}.
Despite the lack of quenched disorders,
the quantum chaos in tensor models seems much more difficult to
analyze by either OTOC or RMT \cite{tensorrev}.
The OTOC and RMT \cite{KAM} may also be used to demonstrate  the quantum chaos in a clean quantum optics model
  called Dicke model which describes the $ N $ qubits interacting with a single photon mode with both rotating wave
  and counter-rotating wave interacting term \cite{u1largen,gold,comment,strongED}.

Gross and Rosenhaus \cite{Gross} generalized the SYK model to a colored SYK,
which contains $ a=1,2, \cdots, f $ colors, each has $ N_a $ sites with  a $ q_a $ body interaction.
The model has a total number of Majorana $ N_t= \sum^{f}_{a=1} N_a $
and a total $ q= \sum^{f}_{a=1} q_a $ body interaction.
It contains $f$ tower of operators.
The SYK model can be treated as the $f=1$ special case with just one tower of operators.
For the balanced case with $N_a=N_t/f$ and $q_a=q/f$,
after the quenched disorder average is performed,
the system has a reduced symmetry $O(N_t/f)\times O(N_t/f) \times\cdots\times O(N_t/f) $,
compared to the SYK model with $N_t=f N_a$ sites
and $q=f q_a$ body interaction which has a full $O(N_t)$ symmetry.
The operator spectrum contains a tower identical to
that of SYK with a $ q= f q_a $ body interaction.
The $ h=2 $ operator which is the lowest dimensional operator in this tower
still leads to the maximal chaos.
There is also a new tower of operators with degeneracy $ f-1 $.
The lowest dimensional operator in this new tower is a $ h=1 $ operator,
whose operator product expansion (OPE) coefficient vanishes.
There maybe some intricate relations between the colored SYK
and the colored tensor (Gurau-Witten) models \cite{graph,tensorSYK1}.
Here, we study the quantum chaos in the colored SYK from RMT
which would be complementary to the OTOC study by Gross and Rosenhaus \cite{Gross}.
For simplicity, we only focus on the balanced $q=4$ cases with $f=2$ and $f=4$ colors.
The analysis is much more involved, the results are dramatically different from the Majorana or complex fermion SYK model.
Our main results are presented in Table III, IV and Fig. 2,5,6.

The 2-colored SYK only shows 3-fold Wigner-Dyson way.
For even $N$ case, there are two conserved parities $(Q_1,Q_2)$ corresponding to the two colors.
We construct one anti-unitary operator $P$
which keep parities and commutes with the Hamiltonian.
For $N \pmod 4=0$, the ELS is in GOE with degeneracy $ d=1 $ in a given parity sector $(Q_1,Q_2)$
and total degeneracy $d_t=1$ in the total parity sector $Q_t=Q_1+Q_2$.
For $N \pmod 4=2$, the ELS is in GUE with degeneracy $ d=1 $ but total degeneracy $ d_t=2 $.
For odd $N$ case, we add two decoupled Majorana fermions with each color at infinity
to construct Hilbert space separately for the two colors.
So it still lead to two conserved parities $(Q_1,Q_2)$ corresponding to the two colors.
Adding two Majorana fermions into the system doubles the Hilbert space,
but also gives one more conserved parity.
There is an additional anti-unitary operator $P_z$
which also keep parities and commutes with the Hamiltonian
and plays complementary roles as $P$.
In both cases of $N \pmod 4=1,3$, the ELS is GOE with $ d=1 $ in a given parity sector$(Q_1,Q_2)$.
But $P$ and $P_z$ exchange their roles in the two cases,
so $d_t=1+1$ in the total parity sector $Q_t=Q_1+Q_2$ in the twice enlarged Hilbert space.
Then exact diagonalizations are performed in the minimal Hilbert space
which match our theoretical classification results (Fig.\ref{twocolorELS}).
We also study a hybrid 2-colored SYK model which violates $(Q_1,Q_2)$ parities,
but conserves the total parity $ Q_t $.
It shows several novel quantum chaotic behaviours at all $N\pmod 4$ values (Fig.3)
which are hidden in the pure 2-colored SYK model.
Our systematic approach can also be extended to a generic case with different $N_a$, $a=1,2$.

The 4-colored case shows dramatic different quantum chaotic behaviours than the 2-colored case.
Due to a possible spectral mirror symmetry,
the 4-colored SYK is classified in the 10-fold way
which may also show non-trivial hard-edge universality.
For the 4-colored case, there are always three independent conserved parities
$(Q_{12}, Q_{23}, Q_{34})$  corresponding to the sum of two of the 4 colors.
For $N$ even case, the three independent conserved parities commute with each other.
We construct one anti-unitary operator $P$ which keep the parities and commutes with the Hamiltonian,
and also find another anti-unitary operator $P_m$ which keep the parities and anti-commutes with the Hamiltonian.
The product of the two anti-unitary operators lead to a unitary chirality operator
$\Lambda=PP_m$ which is nothing but the individual parity of each color,
which anti-commutes with the Hamiltonian.
So when $N\pmod 4=0,2$, the ELS is in class BDI and CI respectively
with $ d=1 $ in a given parity sector $(Q_{12},Q_{23},Q_{34})$.
In addition to the bulk RMT index $ \beta=1 $, due to the chiral (mirror) symmetry,
they also have the edge exponent $ \alpha=0,1 $ respectively.
It is the chiral symmetry which dictates such a non-trivial ``bulk-edge'' correspondence.
For $N$ odd case, $Q^{0}_{23}$ does not commute with the other two parities anymore.
So one only have two mutually conserved parities  $(Q^{0}_{12}, Q^{0}_{34})$.
Then we add four decoupled Majorana fermions with each color at infinity
to construct Hilbert space separately for the four colors.
This enlarges the Hilbert space by $4$ times,
but also lead to two more conserved parities.
So the complete set of mutually commuting conserved parities
becomes $(Q_{12},Q_{23},Q_{34},Q_{0t})$ in the enlarged Hilbert space.
The two anti-unitary $P,P_m$ and the chirality operator $\Lambda$
still exist after shifting $N \to N+1$,
but $P$ keep the parities and $P_m$ swaps each the parities.
When $N\pmod 4=3,1$, the ELS is in Class AI in both cases with $ d=1 $.
We also identify another anti-unitary operator $P_z$ which commutes with the Hamiltonian,
but it swaps each party in $(Q_{12}, Q_{23}, Q_{34})$,
thus keep the same total parity $ Q_t= Q_{12}+ Q_{34} $ and $ Q_{0t} $.
This fact leads to $ d_t=2 $ in the total parity $ ( Q_t, Q_{0t} ) $ when $ N $ is odd.
Then exact diagonalizations are performed to confirm our theoretical results
(Fig.\ref{fourcolorELS}), especially the edge exponent for $N\pmod 4=0,2$ (Fig.\ref{fourcoloredge}).
We also study a hybrid 4-colored SYK model which violates $(Q_{12},Q_{23},Q_{34})$ parities,
but conserves the total parity $ (Q_t, Q_{0t} ) $.
It shows several novel quantum chaotic behaviours at all $N\pmod 4$ values (Fig.3)
which are hidden in the pure 4 colors SYK model.
Our systematic approach can also be extended to the imbalanced cases
with different $N_a$, $a=1,2,3,4$.
The broad impacts of the methods and results achieved in the paper and some perspectives are summarized in the conclusion section.

As a by-product, we develop a systematic RMT to study the energy level statistics of 2 or 4 un-correlated random matrix ensembles.
the apply the new RMT to study several salient features of the 2- or 4-colored hybrid SYK models in Sec.IV and VI respectively.

Finally, in the three appendices, we discuss the inter-color representation
which is independent of $N$ is odd or even,
so can be most conveniently used to perform our exact diagonalizations in the minimum Hilbert space.
We perform our classifications on 2- and 4-colored SYK models
and their corresponding hybrids in this minimum Hilbert space.
To be compared to the results achieved with intra-color representations in the main text,
when $ N $ is odd, we also add decoupled Majorana fermions at $\infty$
and perform our classifications in the enlarged Hilbert space.
We reach the same conclusions among the three different classification schemes
(namely, the two different inter-color scheme in this appendix
and the intra-color scheme in the main text)
which may bring additional and considerable insights into the physical picture.

%
\section{Energy level statistics in pure and mixed Random Matrix Theory}
\label{sec:RMT}
In this section, we first review the known results on the statistics of the nearest neighbour
(NN) energy level spacings initiated in \cite{NNr} and
next-nearest neighbor (NNN) energy level spacings initiated in Ref.\cite{KAMSYK}.
in a pure random matrix ensembles.
Then we generalize the NN and NNN statistics to the case with mixed 2 and 4
un-correlated and identically distributed (UCID) random matrix ensembles.
The results will be heavily used in the following sections when discussing
2 color or 4 color hybrid SYK models.

\subsection{The statistics of NN and NNN energy level spacings in  pure random matrix ensembles }

Let $\{ e_n \}$ be an ordered set of eigen-energy obtained from Hamiltonian,
then the energy level spacing is $s_n=e_{n+1}-e_{n}$
and the ratios of NN energy level spacings and NNN energy level spacings
are defined as $r_n=s_{n+1}/s_{n}$ and $r'_n=(s_{n+3}+s_{n+2})/(s_{n+1}+s_{n})$, respectively.

By considering a $2\times2$ matrices system,
Wigner derived a simple approximate probability distribution function (Wigner surmise):
$P_{w,\beta}(s)=a_\beta s^\beta e^{-b_\beta s^2}$,
where $\beta=1,2,4$ is the Dyson index for
the Gaussian Orthogonal Ensemble (GOE),
the Gaussian Unitary Ensemble (GUE),
and the Gaussian Symplectic Ensemble (GSE) respectively.
The probability distribution function for independent random energy levels
yield Poisson distribution $P_{p}(s)=e^{-s}$.
However, in order to compare different results from different systems,
the energy levels will need an unfolding procedure,
which is not convenient when large enough statistics is not available.

The NN ratio $r$ and NNN ratio $r'$
are introduced  respectively in \cite{NNr} and \cite{KAMSYK} to overcome the difficulties in unfolding.
Because taking the two ratios can get rid of the dependence on the local density of states, so the
unfolding becomes unnecessary.
By considering $3\times3$ matrices system, the authors in \cite{NNr}
obtained the Wigner-like surmises of
the ratio of NN level spacings distribution
$P_w(r)=\frac{1}{Z_{\beta}}\frac{(r+r^2)^{\beta}}{(1+r+r^2)^{1+3\beta/2}}$,
where $\beta=1,2,4$
and $Z_{\beta}=8/27, 4\pi/81\sqrt{3},  4\pi/729\sqrt{3}$ for GOE, GUE and GSE respectively.
The Poisson result is $P_p(r)=\frac{1}{(1+r)^2}$.
The distribution function $P_W(r)$ has the same level repulsion at small $r$ as $P_W(s)$,
namely, $P_W(r)\sim r^\beta$.
However, the large $r$ asymptotic behavior $P_W(r)\sim r^{-(2+\beta)}$ is
dramatically different than the fast exponential decay of $P_W(s)$.

By considering a $5\times5$ matrices system, a Wigner-like surmises
of the ratio of NNN level spacings distribution was obtained by us in Ref.\cite{KAMSYK}.
The asymptotic behavior of $P^{(2)}_{w}(r')$ is different from $P_{w}(r)$:
$P^{(2)}_{w}(r')\sim {r}^{\prime3\beta+1}$ when $r'$ is small,
and $P^{(2)}_{w}(r')\sim {r}^{\prime -3(\beta+1)}$ when $r'$ is large.
The Poisson result is $P^{(2)}_p(r')=\frac{6r'}{(1+r')^4}$.
Instead of the lengthy analytical results from Wigner-like surmises detailed in \cite{KAMSYK},
an approximate, but precise and useful relation between the probability distribution function
of NN ratio and NNN ratio is found as $P^{(2)}_{w,\beta}(r')\approx P_{w,3\beta+1}(r)$ by equating $r=r'$.

It was known the NN ratio satisfies the functional equation
$P(r)=\frac{1}{r^2}P(\frac{1}{r})$, so does the
NNN ratio after replacing $r$ with $r'$.
The property enable us to restrict the study to the range $[0,1]$
by considering the variable $\tilde{r}=\min\{r,1/r\}$ and $\tilde{r}'=\min\{r',1/r'\}$.
Thus above surmise yields an analytic expression for the mean values
$\langle \tilde{r}\rangle_w=0.386,0.536,0.603,0.676$
and $\langle \tilde{r}'\rangle_w=0.500,0.677,0.734,0.791$
for Poisson, GOE, GUE, GSE, respectively.
From these mean values, one can also find
$\langle \tilde{r}\rangle_\text{GSE}\approx\langle \tilde{r}'\rangle_\text{GOE}$
which is just a special case of $\beta\to 3\beta+1$ rule with $\beta=1$.

The motivation to introduce $r'$ in \cite{KAMSYK} is
to deal with the case with nearly double degenerate energy levels.
When the unperturbed Hamiltonian has double degenerate levels, then
a small perturbation will lead to nearly double degenerate levels.
Then $\langle \tilde{r}\rangle$ can be very close to zero
and rapid changes as the perturbation is increased.
However, $\langle \tilde{r}'\rangle$ may remain unchanged,
so it becomes a much better criterion to characterize the quantum chaos in this regime.
Note that $\langle \tilde{r}\rangle$ and $\langle \tilde{r}'\rangle$
are not expected to satisfy theirs values listed in Table I in \cite{KAMSYK}  when the energy levels are nearly doubly  degenerate.
Namely, when $\langle \tilde{r}\rangle$ is close to zero,
$\langle \tilde{r}'\rangle$ will be close to $0.386,0.536,0.603,0.676$.
This  suggests that if one split all energy levels into two sets, one set of energy levels satisfy
Poisson, GOE, GUE, GSE, respectively,
For this reason which was detailed in \cite{KAMSYK}, both $\langle \tilde{r}\rangle$ and $\langle \tilde{r}'\rangle$
will be evaluated throughout this paper.

\subsection{The ELS of mixed 2 or 4 mixed random matrix ensembles}

In one case, the unperturbed Hamiltonian has a double degeneracy dictated by a symmetry,
then a small perturbation breaks the symmetry, then the double degeneracy into two nearly degenerated levels.
This case was discussed in the last subsection.
In another case,  the unperturbed Hamiltonian do not have any level degeneracy
then a small perturbation just breaks some symmetry of the  Hamiltonian, then it
will mix different sets of energy levels
which can be labeled by different conserved quantities in the absence of the perturbation.
For an infinitesimal perturbation, there is still an obstacle to collect statistics for
individual sets of energy levels.
In this subsection, we investigate the ELS of mixed 2 or 4 un-correlated and
identically distributed ( UCID ) random matrix ensembles \cite{also}.

We begin with mixed 2 UCID random matrix ensembles with size $N=2$,
of which the joint probability distribution function can be written as
\begin{align}
    &p(e_1,e_2,e_3,e_4)
	=p(e_1,e_2)p(e_3,e_4) \nonumber\\
	&\propto
	    e^{-\frac{1}{2}(e_1^2+e_2^2)}|e_1-e_2|^\beta
	    e^{-\frac{1}{2}(e_3^2+e_4^2)}|e_3-e_4|^\beta\>,
\end{align}
where the level repulsion only exists between $e_1$ and $e_2$, $e_3$ and $e_4$.

The ordering of levels can be summarized in following 3 independent cases:
i) level ordering $e_1\leq e_2\leq e_3\leq e_4$,
and two NN ratios defined as $(e_3-e_2)/(e_2-e_1)$ and $(e_4-e_3)/(e_3-e_2)$;
ii) level ordering $e_1\leq e_3\leq e_2\leq e_4$,
and two NN ratios defined as $(e_2-e_3)/(e_3-e_1)$ and $(e_4-e_2)/(e_2-e_3)$;
iii) level ordering $e_1\leq e_3\leq e_4\leq e_2$,
and two NN ratios defined as $(e_4-e_3)/(e_3-e_1)$ and $(e_2-e_4)/(e_4-e_3)$.
Other orderings can be related to these 3 ones by taking the full advantage of symmetries in
$p(e_1,e_2)=p(e_2,e_1)$ and $p(e_3,e_4)=p(e_4,e_3)$.

Now we generalize the $r$-statistics defined in Ref.\cite{jpdf-r}
to mixed-2 random matrix ensembles,
the probability density function of $r$ can be calculated from
\begin{align}
    P_\text{mix-2}(r)
	\propto&\int_{e_1\leq e_2\leq e_3\leq e_4} \prod_{i=1}^{4}de_i
		p(e_1,e_2,e_3,e_4) \nonumber\\
			&\times\Big
			[\delta\Big(r-\frac{e_3-e_2}{e_2-e_1}\Big)
			+\delta\Big(r-\frac{e_4-e_3}{e_3-e_2}\Big)\Big]	\nonumber\\
	+&\int_{e_1\leq e_3\leq e_2\leq e_4} \prod_{i=1}^{4}de_i
		p(e_1,e_2,e_3,e_4) \nonumber\\
			&\times\Big
			[\delta\Big(r-\frac{e_2-e_3}{e_3-e_1}\Big)
			+\delta\Big(r-\frac{e_4-e_2}{e_2-e_3}\Big)\Big]	\nonumber\\
	+&\int_{e_1\leq e_3\leq e_4\leq e_2} \prod_{i=1}^{4}de_i
		p(e_1,e_2,e_3,e_4) \nonumber\\
			&\times\Big
			[\delta\Big(r-\frac{e_4-e_3}{e_3-e_1}\Big)
			+\delta\Big(r-\frac{e_2-e_4}{e_4-e_3}\Big)\Big]\>,
\label{eq:int}
\end{align}
where only 3 independent cases are considered
and the other cases only contribute an overall factor.

After changing variables, the integral Eq.\eqref{eq:int} can be rewritten in a neat form
\begin{align}
    P_\text{mix-2}(r)\propto\int \prod_{i=1}^{4}de_i
	    p_s(e_1,e_2,e_3,e_4)\sum_{j=1}^2\delta(r-r_j)
\label{eq:mix-2}
\end{align}
where the NN ratio $r_j=(e_{j+2}-e_{j+1})/(e_{j+1}-e_{j})$ and
$p_s(e_1,e_2,e_3,e_4)
=[p(e_1,e_2,e_3,e_4)+p(e_1,e_3,e_2,\allowbreak e_4)+p(e_1,e_3,e_4,e_2)]/3$
is fully symmetrized in $\{e_n\}$.

The integrals in Eq. \eqref{eq:mix-2} can be evaluated analytically,
and the results are
\begin{align}
    P_\text{mix-2}(r)
	&=A_\beta(r)+\frac{1}{r^2}A_\beta(1/r)
\end{align}
where $A_{\beta}(r)$ ($\beta=1,2,4$) are
$A_1(r)=(1+r)[15+31r+34r^2+16r^3-(3+r-4r^3)\sqrt{3(3+4r+4r^2)}]
/[4(1+r+r^2)^2(3+4r+4r^2)^{3/2}]$,
$A_2(r)=3[\sqrt{3}-(21+70r+92r^2+88r^3+40r^4+16r^5)(3+4r+4r^2)^{-5/2}]
/[2\pi(1+r+r^2)]$, and
$A_4(r)=[11+33r+39r^2+23r^3+39r^4+33r^5+11r^6-3\sqrt{3}(1+2r)(279+1953r+7467r^2+18731r^3+33883r^4+45581r^5+46551r^6+36224r^7+22172r^8+11736r^9+6712r^{10}+4128r^{11}+2144r^{12}+672r^{13}+96r^{14})
(3+4r+4r^2)^{-9/2}]/[2\sqrt{3}\pi(1+r+r^2)^4]$.

To investigate $r'$-statistic of mixed two random matrix ensembles,
one need consider size $N=3$ case, and
the joint probability distribution function can be written as
\begin{align}
    &p(e_1,e_2,e_3,e_4,e_5,e_6)
	=p(e_1,e_2,e_3)p(e_4,e_5,e_6) \nonumber\\
	&\propto
	    e^{-\frac{1}{2}(e_1^2+e_2^2+e_3^2)}|(e_1-e_2)(e_1-e_3)(e_2-e_3)|^\beta \nonumber\\
	   &\times e^{-\frac{1}{2}(e_4^2+e_5^2+e_6^2)}|(e_4-e_5)(e_4-e_6)(e_5-e_6)|^\beta\>.
\end{align}
After considering all the possible energy level orderings,
the probability density function $P_\text{mix-2}(r')$
can be expressed as
\begin{align}
    P_\text{mix-2}(r')
	\propto\int \prod_{i=1}^{6} de_i
		p_s(e_1,\cdots,e_6)\sum_{j=1}^{2}\delta(r-r'_j)
\end{align}
where the NNN ratio $r'_j=(e_{j+4}-e_{j+2})/(e_{j+2}-e_{j})$ and
$p_s(e_1,e_2,e_3,e_4,e_5,e_6)
=[p(e_1,e_2,e_3,e_4,e_5,e_6)+p(e_1,e_2,e_4,e_3,e_5,e_6)
+p(e_1,e_3,e_4,e_2,e_5,e_6)+\cdots]/10$
is fully symmetrized in $\{e_n\}$.

Just like the calculation of $P_\text{mix-2}(r)$,
the $P_\text{mix-2}(r')$ can be evaluated exactly.
Although the analytical result is lengthy,
the numerical evaluation of the integration is rather easy.
Similar to the discussion in Sec. II A,
$\tilde{r}=\min\{r,1/r\}$ and $\tilde{r}'=\min\{r',1/r'\}$ can be defined in the mixed case also.
The mean values of $\langle \tilde{r}\rangle$ and $\langle \tilde{r}'\rangle$
are listed in Table I. These results will be used to distinguish
the chaos regime from the integrable regime for the 2- colored hybrid SYK models to be discussed in Sec.IV.

\renewcommand{\arraystretch}{1.4}
\renewcommand\tabcolsep{8pt}
\begin{table}[!htb]
\caption{
List of numerical values of averages $\langle \tilde{r}\rangle$ and $\langle \tilde{r}'\rangle$
for the mixed 2 UCID random matrix ensembles.
The values of $\langle \tilde{r}\rangle_W$ and $\langle \tilde{r}'\rangle_W$
are calculated from the derived $N=2$ surmise and $N=3$ surmise, respectively.
The values of $\langle \tilde{r}\rangle_\text{num}$ and $\langle \tilde{r}'\rangle_\text{num}$
are calculated from diagonalizing the corresponding mixed-2 GOE
(real), GUE (complex) and GSE (quaternion) matrices of size $N=1000$
with Gaussian distributed entries, averaged over $10^5$ histograms.
It is interesting to see they are slightly above the corresponding Poisson values:
$\langle\tilde{r}\rangle_P\approx0.386$ and $\langle\tilde{r}'\rangle_P\approx0.5$.  }
\begin{tabular}{ c|ccc }
\toprule
mix-2					&$\beta=1$  &$\beta=2$	&$\beta=4$	\\ \hline
$\langle\tilde{r}\rangle_W$		&0.423	    &0.421	&0.394		\\
$\langle\tilde{r}'\rangle_W$		&0.600	    &0.649	&0.709		\\ \hline
$\langle\tilde{r}\rangle_\text{num}$	&0.424	    &0.422	&0.410		\\
$\langle\tilde{r}'\rangle_\text{num}$	&0.599	    &0.650	&0.706		\\ \hline
\end{tabular}
\label{tab:mix-2}
\end{table}

In the same manner, the ELS of
mixed-4 UCID random matrix ensembles with $N=2$ can be studied.
We skip the analytical calculations for probability density function,
and only list the numerical results of
$\langle \tilde{r}\rangle$ and $\langle \tilde{r}'\rangle$ in Table II.
In all the cases of $ \beta=1,2,4$, we obtain $\langle\tilde{r}\rangle\approx0.39$
and $\langle\tilde{r}'\rangle\approx 0.535$ which are quite in-sensitive to the values of $ \beta $,
so the differences between mixed 4 GOE/GUE/GSE are almost washed away.
They are also different from those in the mixed 2 UCID random matrix ensembles listed in the Table I.
When comparing with the Poisson results
$\langle\tilde{r}\rangle_P\approx0.386$ and $\langle\tilde{r}'\rangle_P\approx0.5$, it is easy to see that
even $\langle\tilde{r}\rangle$ value is very close to the Poisson result,
$\langle\tilde{r}'\rangle$ is still easily distinguishable from the Poisson's value ( in fact, quite close to
the GOE value of $\langle\tilde{r}\rangle=0.5359 $ ).
This fact will be used to distinguish
the chaos regime from the integrable regime for the 4- colored hybrid SYK models to be discussed in Sec.VI.

\begin{table}[!htb]
\caption{The same as Table \ref{tab:mix-2},
but r-statistic for the mixed 4 UCID random matrix ensemble. They seem  quite in-sensitive to the values of $ \beta $.
They are closer to the corresponding Poisson values than the 2-mixed case listed in Table I.
}
\begin{tabular}{ c|ccc }
\toprule
mix-4					&$\beta=1$	&$\beta=2$	&$\beta=4$  \\ \hline
$\langle\tilde{r}\rangle_\text{num}$	&0.396		&0.393		&0.389	    \\
$\langle\tilde{r}'\rangle_\text{num}$	&0.535		&0.537		&0.533	    \\ \hline
\end{tabular}
\label{tab:mix-4}
\end{table}

%
\section{The 2-colored $ q=4 $  SYK}

The 2-colored $ a=1,2 $ SYK with $q_1=q_2=2$ and $N_1=N_2=N$
can be written as
\begin{equation}
  H_{1122}=\sum^{N}_{i<j;k<l} J_{ij;kl} \chi_{1i} \chi_{1j} \chi_{2k} \chi_{2l}
\label{gr2}
\end{equation}
where  $J_{ij;kl}=-J_{ji;kl}=-J_{ij;lk}$ are real and satisfy the Gaussian distribution with
mean value $ \langle J_{ij;kl} \rangle=0$
and variance $\langle J^2_{ij;kl} \rangle= 2J^2/N^3$.

\begin{figure}[!htb]
\centering
\includegraphics[width=0.9\linewidth]{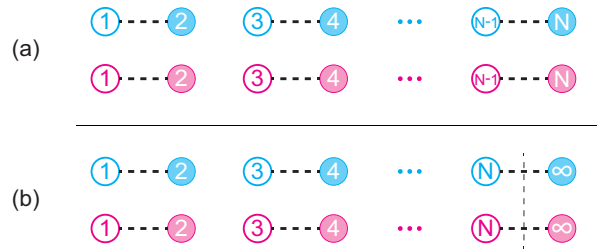}
\caption{
The 2-colored SYK with (a) $N$ even and (b) $N$ odd.
Majorana fermions are represented by dots.
Each dot has a associated number and color,
representing its site index and color index respectively.
The solid dot means the Majorana is represented by a real matrix,
while the empty dot means the Majorana is represented by a imaginary matrix.
The dashed line connecting them means a complex fermion.
In (b), the long vertical dashed line separate the system
from two decoupled Majorana added at infinity.
The same conventions apply to the other figures. }
\label{twocolor.eps}
\end{figure}

At the first sight, no matter $N$ is even or odd,
one can always introduce $N$ complex fermions by combining the two colors
$c_i= (\chi_{1i}-i\chi_{2i})/\sqrt{2}$ and
$c^{\dagger}_i=(\chi_{1i}+i \chi_{2i})/\sqrt{2}$.
Both $c_i$ and $c^{\dagger}_i$ can be represented by real matrices
and the particle-hole symmetry operator can be defined as
$P_{12}=K\prod^{N}_{i=1}(c^{\dagger}_i+c_i)$,
where $K$ is complex conjugate operator.
This way of pairing Majorana fermions with different colors to form complex fermions
is called inter-color scheme.
As to be shown in the appendix A, this construction using $P_{12}$ across the two
colors is an alternative representation to discuss the symmetry class of the Hamiltonian.
However, we choose a different approach called intra-color scheme in the main text.
Both approaches have their own advantages, and they are complementary to each other.

Due to the absent of spectral mirror symmetry,
2-colored SYK models will be classified in 3-fold way.
The 3-fold way classifies 3 Wigner-Dyson ensembles: GUE, GOE, and GSE.
The classification need an anti-unitary operator $T_+$,
which commutes with the Hamiltonian as well as all compatible conserved quantities.
If such a $T_+$ exist,
then its squared value $(T_+)^2=+1$ means the Hamiltonian is GOE
and  its squared value $(T_+)^2=-1$ means the Hamiltonian is GSE.
If such a $T_+$ do not exist, the Hamiltonian must be GUE.

\subsection{$ N $ even case }

In $N$ even case, just following Ref.\cite{KAMSYK},
the intra-color scheme need split the site $i$ into even and odd sites
(see Fig.\ref{twocolor.eps}(a)),
then introduce $ N_c=N/2 $ complex fermions for each color
$ c_{1i}= ( \chi_{1,2i}-i \chi_{1,2i-1} )/\sqrt{2}$
and $c^{\dagger}_{1i}= ( \chi_{1,2i}+i \chi_{1,2i-1} )/\sqrt{2} $.
The particle-hole symmetry operator can be defined as
$P_1=K\prod^{N_c}_{i=1} ( c^{\dagger}_{1i} + c_{1i} ) $
or $R_1=P_1(-1)^{Q_1} = K \prod^{N_c}_{i=1} ( c^{\dagger}_{1i} -c_{1i} ) $,
but using $P_1$ is enough for the symmetry classification
and $R_1$ will not lead to any new result.
It is easy to show
$P_1 c_{1i} P_1= \eta c^{\dagger}_{1i}$,
$P_1 c^{\dagger}_{1i} P_1= \eta c_{1i}$,
$P_1 \chi_{1i} P_1= \eta \chi_{1i} $,
and $ P_1^2= (-1)^{\lfloor N_c/2 \rfloor} $,
where $ \eta=(-1)^{\lfloor (N_c-1)/2 \rfloor} $.
The number operator of color-1 fermions $ Q_1= \sum^{N_c}_{i=1} c^{\dagger}_{1i} c_{1i} $
is not a conserved quantity, but its parity $ (-1)^{Q_1} $ commutes with $ H_{1122} $.
The fact $P_1Q_1P^{-1}_1= N_c-Q_1 $ also justifies $P_1$
as an anti-unitary particle-hole transformation.
One can similarly construct $ P_2 $ operator from color-2 fermions,
so it is convenient to characterize the Hilbert space in terms of
the conserved joint parities $(Q_1,Q_2)$ which block diagonalize it into 4 sectors:
$(Q_1,Q_2)=$(Even,Even), $(Q_1,Q_2)=$(Even,Odd),
$(Q_1,Q_2)=$(Odd,Even), and $(Q_1,Q_2)=$(Odd,Odd).
Unfortunately, neither $ P_1 $ nor $ P_2 $,
neither commute nor anti-commute with the Hamiltonian,
but $ P_1, P_2, R_1, R_2 $
can be used as building blocks to construct operators which will do the job.

To construct an anti-unitary operator which commute nor anti-commute with the $H_{1122}$,
we introduce
\begin{align}
    P=K\prod^{N_c}_{i=1} (c^{\dagger}_{1i}+c_{1i})(c^{\dagger}_{2i}+c_{2i})=KP_1P_2
\label{p1122}
\end{align}
which can be contrasted to the similar operator in the 4-colored case Eq.\eqref{p1234c}
to be discussed in the Sec. IV.
One can show that:
\begin{equation}
    P\chi_{ai}P
	=(-1)^{\lfloor \frac{N_c}{2}\rfloor }\eta\chi_{ai}
	=(-1)^{N_c-1} \chi_{ai}, \quad a=1,2
\label{p1122act}
\end{equation}
It is easy to see $ P^2=(-1)^{N_c} $,
$ P Q_1 P^{-1}= N_c-Q_1 $, $ P Q_2 P^{-1}= N_c-Q_2 $,
and the operator $ P $ indeed commutes with the Hamiltonian $ [P,H_{1122} ]=0 $.
Other combinations of $P_1$, $P_2$, $R_1$, $R_2$ play exact the same role as $P$,
thus no anti-commuting operators exist and spectral mirror symmetry is absent.
This is the main difference from the 4-colored case to be discussed in the Sec. IV,
where one can find two anti-unitary operators, one $P$ in Eq.\eqref{p1234c} commutes,
another $P_{m}$ in Eq.\eqref{pm1234c} anti-commutes with the Hamiltonian.

\begin{figure}[!htb]
\centering
	\includegraphics[width=\linewidth]{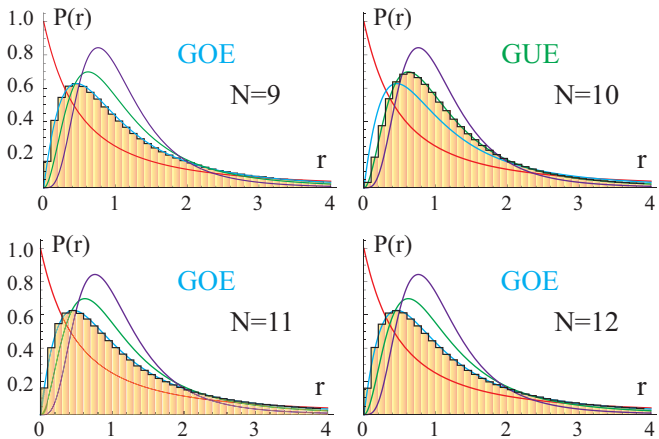}
	\caption{Distribution of the ratio of consecutive level spacings $P(r)$
	for 2-colored SYK with various $N=9,10,11,12$.
	When $N\pmod 4=1,2,3,0$,
	the energy level statistics show GOE, GUE, GOE, GOE respectively,
	which agree with symmetry analysis summarized in Table III.
	The 4 background curves are $P(r)$ of Poisson (red), GOE (blue), GUE (green), GSE (purple),
	respectively.}
\label{twocolorELS}
\end{figure}

For $N \pmod 4=0$, $ N_c=N/2 $ is even,
then $P$ maps $(Q_1, Q_2)$ sector to the same joint parities sector and $P^2=1$,
the ELS is GOE.
The level degeneracy is $ d=1 $ in a given $(Q_1,Q_2) $ sector.
Because the 4 sectors are un-related,
the level degeneracy is $ d_t=1 $ in total parity sector $ Q_t= Q_1 + Q_2 $.

For $N \pmod 4=2$, $ N_c=N/2 $ is odd, 
then $P$ maps $(Q_1,Q_2)$ sector to a different joint parities sector with $(Q_1+1, Q_2+1)$,
the ELS is GUE.
The level degeneracy is $ d=1 $ in a given $(Q_1,Q_2) $ sector.
However, if just focus on the total parity $Q_t$,
it is still mapped to the same total parity sector,
so it has the $ d_t=2 $ double degeneracy  \cite{comp2} in a given total parity sector $ Q_t $.
Note the 4 sectors can still be separated into two sectors with a given total parity $ Q_t $,
which maybe useful when we consider a quadratic perturbation such as Eq.\eqref{gr2hyv}
which violates the separate parities $(Q_1,Q_2)$, but still conserve the total parity $Q_t$.

\subsection{$ N $ odd case}

When $N \pmod 4=1,3$, the above procedures for even $ N $ needs to be modified.
In fact, one can still take the advantage of the above representation with $N$ even case by
adding two decoupled Majorana fermions
$\chi_{1,N+1}=\chi_{1,\infty}$ and $\chi_{2,N+1} = \chi_{2,\infty}$
to make the parity conservation in the color 1 and color 2 respectively and explicitly
(see Fig.\ref{twocolor.eps}(b)).
In doing so, one also doubles the Hilbert space
when comparing with the Hilbert space without $\chi_{1,\infty}$ and $\chi_{2,\infty}$.
Similar strategy was used before to study the symmetry protected topological phase of
odd number of Majorana chain \cite{kitSPT} and the ELS of the SYK model with $N$ odd \cite{MBLSPT}.
Then one can still define $P_1$, $P_2$, and $P$ with $ N_c=(N+1)/2$ as before
and Eq.\eqref{p1122act} still applies.
From the fact that $\chi_{1\infty} $ and $\chi_{2\infty} $ do not appear in the Hamiltonian,
we have two more building blocks:
$Z_1=P_1\chi_{1 \infty}= K\prod^{N_c-1}_{i=1}(c^{\dagger}_{1i}+c_{1i})$
and $Z_2=P_2 \chi_{2 \infty}= K \prod^{N_c-1}_{i=1} (c^{\dagger}_{2i}+c_{2i})$,
which can be obtained by factoring out
$\chi_{1\infty}$ and $\chi_{2\infty}$ from $P_1$ and $P_2$ respectively.
So when $N$ is odd,  $P_1$, $P_2$, $R_1$, $R_2$ and $Z_1$, $Z_2$
are building blocks to construct various operators.

In addition to $P$ operator introduced in Eq.\eqref{p1122},
another special operator can be constructed for $N$ odd case:
\begin{equation}
    P_z=K\prod^{N_c-1}_{i=1}
	( c^{\dagger}_{1i} + c_{1i} ) ( c^{\dagger}_{2i} + c_{2i} )
	= K Z_1 Z_2
\label{z1z2}
\end{equation}
One can show that:
\begin{equation}
    P_z \chi_{ai} P_z
	=  -(-1)^{ \lfloor \frac{N_c}{2}\rfloor } \eta \chi_{ai}=(-1)^{N_c} \chi_{ai},~ a=1,2
\label{z1z2act}
\end{equation}
where, of course, as usual,  $ i=\infty $ is always excluded.
It is also easy to see that $ P^2_z=(-1)^{N_c-1} $
and $P_z$ still commutes with the Hamiltonian $ [P_z,H_{1122}]=0 $.
It also leads to $ P_z Q_a P^{-1}_z= N_c-1-Q_a + 2 n_{a\infty} $,
where $ n_{a \infty}= c^{\dagger}_{a\infty} c_{a\infty}=
1/2- i \chi_{a\infty} \chi_{a,N}$ and $a=1,2$.

When $N \pmod 4=3$, $N_c$ is even,
the $P$ operator maps $(Q_1,Q_2)$ to a sector with the same joint parities and $P^2=1$.
So the ELS is in GOE, and the level degeneracy $d=1$ at a given parities sector $(Q_1,Q_2)$.
When using the $ P_z $ operator in Eq.\ref{z1z2}
which maps $(Q_1,Q_2)$ to a sector with opposite joint parities $(Q_1+1,Q_2+1)$,
one can conclude the double degeneracy $ d_t=1+1 $ in a given total parity sector $ Q_t $.

When $N \pmod 4=1$, $N_c$ is odd,
the $P_z$ operator maps $(Q_1,Q_2)$ to a sector with the same joint parities and $P^2_z=1$.
So the ELS is still in GOE, and the level degeneracy $d=1$ at a given parities sector $(Q_1, Q_2)$.
While the $P$ operator maps $(Q_1,Q_2)$ to a sector with opposite parities $(Q_1+1,Q_2+1)$,
one can conclude the double degeneracy $ d_t=1+1 $ in a given total parity sector $ Q_t $.

In summary, when $N$ is even, there are two cases:
when $N \pmod 4=0$, the ELS is in GOE;
when $N \pmod 4=2$, the ELS is in GUE.
One use the $P$ operator Eq.\eqref{p1122} in both cases.
When $N$ is odd, one may need to add decoupled Majorana fermion at infinity for each color.
There are also two cases, both cases are GOE.
When $N \pmod 4=3$, one still use the $P$ operator Eq.\eqref{p1122},
but when $N \pmod 4=1$, one must use the $P_z$ operator Eq.\eqref{z1z2}.
These theoretical results are listed in the Table III and
are confirmed by the exact diagonalizations shown in Fig.\ref{twocolorELS}.

\renewcommand{\arraystretch}{1.5}
\renewcommand\tabcolsep{5.5pt}
\begin{table}
\label{tab:2c}
\caption{The ELS and degeneracy of the 2-colored SYK model.
The degeneracy $d=1$ is at a given parity sector $(Q_1,Q_2)$.
The total degeneracy $d_t$ is at a total parity sector $Q_t=Q_1+Q_2$.
(a) $N$ even case.
When $N \pmod 4=0$, $P$ maps $ (Q_1,Q_2) $ to itself.
When $N \pmod 4=2$, $P$ maps $ (Q_1,Q_2) $ into $ (Q_1+1,Q_2+1) $.
(b) $N$ odd case.
When $N \pmod 4=1$, $P_z$ maps $ (Q_1,Q_2) $ to itself, $ P $ maps $ (Q_1,Q_2) $ into $ (Q_1+1,Q_2+1) $.
When $N \pmod 4=3$, $P$ maps $ (Q_1,Q_2) $ to itself, $ P_z $ maps $ (Q_1,Q_2) $ into $ (Q_1+1,Q_2+1) $.
So $ P $ and $ P_z $ exchange their roles in the two cases of odd $ N $.
So $ d_t $ is the degeneracy in the enlarged Hilbert space
which may not be seen in the exact diagonalization doing in the minimum Hilbert space,
which is defined without adding the two Majorana fermions at $\infty$.
When doing exact diagonalization in the minimum Hilbert space,
only the $d_t=2$ at $N \pmod 4=2$ case can be seen as shown in Fig.\ref{color1122hy} (a) and (b).
However, the $ d_t=1+1 $ at $N \pmod 4=1,3$ cases can not be seen (see also Appendix A).  }
\centering
\begin{tabular}{c|c|c|c|c}
\toprule
  $N\pmod 4$	& 0	    & 1		& 2	    & 3		\\  \hline
  ELS		& GOE	    & GOE	& GUE	    & GOE	\\  \hline
  $ \beta $	& 1	    & 1		& 2	    & 1		\\  \hline
  $(Q_1,Q_2) $  & $d=1$	    & $d=1$	& $d=1$	    & $d=1$     \\  \hline
  $Q_t=Q_1+Q_2$ & $d_t=1$   & $d_t=1+1$ & $d_t=2$   & $d_t=1+1$   \\   \hline
\end{tabular}
\end{table}

\section{hybrid 2-colored $ q=2 $ and $ q=4 $ SYK model. }

In the following, we will discuss a parity $(Q_1,Q_{2})$ violating hybrid 2-colored SYK model
Eq.\eqref{gr2hyv}, which still conserves the total parity $Q_t$.
It can be used to study the stability of quantum chaos
and Kolmogorov-Arnold-Moser theorem in the $f=2$ colored SYK models \cite{KAMSYK}.
Furthermore, one can demonstrate the importance of identifying the maximal symmetry,
the largest conserved quantities and the minimal Hilbert space to
do the correct classifications in the RMT.
A small perturbation $ K/J \rightarrow 0 $ limit which breaks $(Q_1,Q_2)$,
but conserves $ Q_t=Q_1+Q_2 $ may also be used to drag out the rich
and novel physics encoded in the Table III from a very effective angle.
This kind of small perturbation may also be used to probe the interior
of a dual black hole in the bulk \cite{pure}.

A 2-colored $ q=2 $ and $q=4$ hybrid SYK model
is hybrid from $H_{1122}$ and $H_{12}$
\begin{equation}
    H^{Hb}_{1122}= \! \sum^{N}_{i<j;k<l}\! J_{ij;kl} \chi_{1i} \chi_{1j} \chi_{2k} \chi_{2l}
	+ i \sum^{N}_{i,j} K_{i;j} \chi_{1i} \chi_{2j}
\label{gr2hyv}
\end{equation}
where $J_{ij;kl}$, $K_{i;j}$ are real and satisfy the Gaussian distribution with
$\langle J_{ij;kl} \rangle=0$, $\langle J^2_{ij;kl} \rangle= 4 J^2/N^3 $ and
$\langle K_{i;j} \rangle=0$, $\langle K^2_{i;j} \rangle= 2 K^2/N $ respectively.
Of course, other hybrid 2-colored  models can be constructed,
but Eq.\eqref{gr2hyv} is the most democratic one between the two colors.
%
For the hybrid model, parities $(Q_1,Q_2)$ do not conserve separately anymore,
but the total parity $Q_t$ remains conserved.
However, $P$ in Eq.\eqref{p1122} (or $Z$ in Eq.\eqref{z1z2} for $N\pmod 4=1$)
not commutes with $H^{Hb}_{1122}$ anymore
due to $\{P,H_{12}\}=0$ (or $\{Z,H_{12}\}=0$),
thus different operators are needed to classify Eq.\eqref{gr2hyv}.

\begin{figure}[!htb]
\centering
\includegraphics[width=\linewidth]{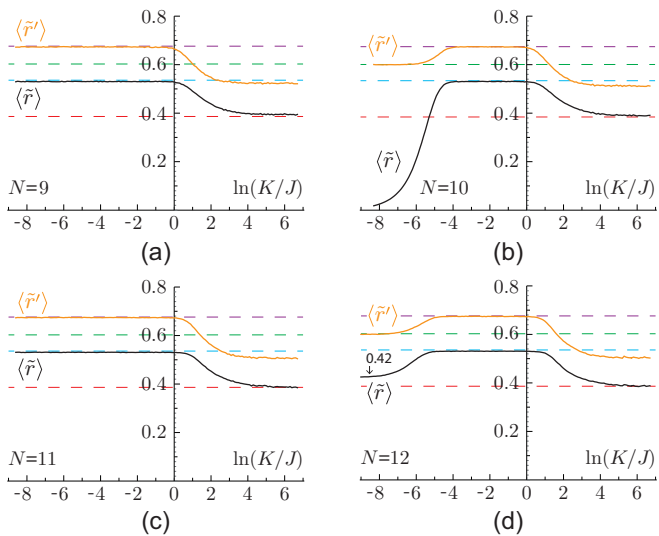}
\caption{(Color online)
The averaged value of the $\tilde{r}$-parameter and $\tilde{r}'$-parameter for
the hybrid 2-colored SYK models with $N=9,10,11,12$.
All the data is taken at a given total parity sector $Q_t=Q_1+Q_2$,
and averaged over $1000$, $800$, $600$, $400$ samples respectively.
Four dashed horizontal lines represent values $0.386$, $0.536$, $0.603$, $0.676$,
and, if there is no mixing or nearly double degeneracy,
$\langle \tilde{r}\rangle$ at those values mean the ELS is Poisson, GOE, GUE, GSE, respectively.
(a) For $ N \pmod 4=1 $, it shows the $ H_{1122} $ (namely, at $ K/J=0 $) is in GOE.
There is a chaotic to non-chaotic transition from the GOE to the Poisson as $ K/J $ increases.
(b) For $ N \pmod 4=2 $, $\langle\tilde{r}\rangle$ (black curve) for NN ELS is rapidly changing,
but $\langle\tilde{r}'\rangle$ (orange curve) for NNN  ELS shows a nice GUE plateau near $K=0$.
(c) For $ N \pmod 4= 3 $, it shows the similar behaviours as that in (a). 
(d) For $ N \pmod 4=0 $, the (slightly above) Poisson-like near $K/J=0$ will be split
into two independent GOE when projected into two separate $(Q_1,Q_2)$.
However, as shown in Sec. II.B, the NN ELS and NNN ELS of the mixed ensemble
of the two un-correlated GOE sectors lead to
$\langle\tilde{r}\rangle\approx0.42$ and $\langle\tilde{r}'\rangle\approx0.6$ listed in Table I.
This is indeed observed here.
All the chaotic to non-chaotic transition
is through the GOE due to the $ P_m $ symmetry at finite $ K/J $.}
\label{color1122hy}
\end{figure}

\subsection{$ N $ even case}

When $N$ is even, $N_c=N/2$,
we can still take advantage of building blocks in the Sec. III
and construct following operator:
\begin{equation}
    P_m= K \prod^{N_c}_{i=1} (c^{\dagger}_{1i}+c_{1i})(c^{\dagger}_{2i}-c_{2i})
	= K P_1 R_2
\label{p1122m}
\end{equation}
then one can show that
\begin{eqnarray}
    P_m \chi_{1i} P_m
	& = & (-1)^{\lfloor \frac{N_c}{2}\rfloor +N_c }\eta \chi_{1i}=-\chi_{1i} ,  \nonumber   \\
    P_m \chi_{2i} P_m
	& = & -(-1)^{\lfloor \frac{N_c}{2}\rfloor+N_c } \eta \chi_{2i}=\chi_{2i} ,
\label{p1122mact}
\end{eqnarray}
Due to the opposite sign in two colors, one can show $[P_m,H_{12}]=0$.
It is obvious $[P_m, H_{1122}]=0$,
thus one can conclude $P_m$ commutes with hybrid Hamiltonian $[P_m,H^{Hb}_{1122}]=0$.
Since $P_m Q_1 P^{-1}_m=N_c-Q_1$ and $P_m Q_2 P^{-1}_m= N_c-Q_2$,
one can show that $P_mQ_tP^{-1}_m=2N_c-Q_t$
and $P_m$ always map to the same total parity sector and $ P^2_m=1 $ always holds.
So surprisingly or counter-intuitively, in sharp contrast to
all the type-I hybrid SYK models studied in \cite{KAMSYK},
the hybrid system is in GOE at a given total parity sector.
This is exactly what is observed in Fig.\ref{color1122hy}.

For $ N \pmod 4=2 $ in Fig.\ref{color1122hy}(b),
it is instructive to look at $ K/J \rightarrow 0 $ limit,
the $ H_{1122} $ at $ K/J=0 $ has two-fold degeneracy $ d_t=2 $
confining to the total parity $ Q_t $ (see Table III).
It consists of two sectors $ (Q_1,Q_2) $ and $ (Q_1+1, Q_2+1) $
which are mapped to each other by the operator $P$ in Eq.\eqref{p1122}.
However, as shown in Fig.\ref{twocolorELS},
when we do the ELS on separate parities $(Q_1,Q_2)$, then the ELS shows GUE.
Indeed, we take just one set of energy levels at any ratio of $ K/J $,
then $\langle \tilde{r}'\rangle$ value tells the set stays at GUE until to $ K/J \sim e^{-6} $.
The other set shows the identical behaviour.
As pointed out in Sec. II,
when NN ratio $\langle \tilde{r}\rangle$ is in its GOE value,
the corresponding NNN ratio $\langle \tilde{r}'\rangle$ would be close to
$\langle \tilde{r}\rangle$'s GSE value;
when $\langle \tilde{r}\rangle$ is in Poisson,
the $\langle \tilde{r}'\rangle$ would be close to $1/2 $.
Fig.\ref{color1122hy}(b) shows
the hybrid 2-colored SYK is in GOE in some range near $ K/J =1 $.
And there is a chaotic to non-chaotic transition
from the GOE to the Poisson as $K/J$ increases.

For $ N \pmod 4=0 $ in Fig.\ref{color1122hy}(d),
in $ K/J \rightarrow 0 $ limit,
the $H_{1122}$ has no degeneracy $ d_t=1 $ (see Table III).
When doing the exact diagonalization in the total parity sector  $ (-1)^{Q_t} $,
the energy levels in two opposite parities sectors $(Q_1,Q_2)$ and $(Q_1+1,Q_2+1)$
are independent of each other and mixed together.
Because there is no level repulsions between the energy levels in the two separate parities sectors,
the ELS may start to show something close to the Poisson \cite{also}.
In Sec. II B, we show that  mixed-2 GOE indeed lead to $\langle \tilde{r}\rangle\approx0.423$
and $\langle \tilde{r}'\rangle\approx0.600$, which agree with the Fig.\ref{color1122hy}(d) precisely.
So the $\langle \tilde{r}\rangle$ and $\langle \tilde{r}'\rangle$
in Fig.\ref{color1122hy}(d) at small $K/J$
indicate  2 un-correlated GOE are mixed together. Then, as shown in Fig.\ref{twocolorELS}(b),
if one do the ELS on separate parities $(Q_1,Q_2)$, then the ELS shows it real face: GOE.
In a given total parity sector, the hybrid 2-colored SYK is in GOE in some range near $ K/J =e^2 $,
there is a chaotic to non-chaotic transition from the GOE to the Poisson as $ K/J $ increases.

\subsection{$ N $ odd case}

When $N$ is odd, as in the $ q=4 $ case discussed in Sec. III B,
after adding $ \chi_{1,N+1} =\chi_{1 \infty} $ and $ \chi_{2,N+1} = \chi_{2 \infty} $,
one can still define $N_c=(N+1)/2$.
Then Eq.\eqref{p1122m} and Eq.\eqref{p1122mact}
still follow and the discussions following them still hold \cite{pmpz}.
So the hybrid system should be in GOE.
In reality, there is a chaotic to non-chaotic transition from GOE to Poisson as $ K/J $ increases.

In order to count the level degeneracy,
one may try the following operator
by replacing $ P_1 $ in Eq.\eqref{p1122m} with $ Z_1 $:
\begin{equation}
    P^{\prime}_m= K \prod^{N_c-1}_{i=1} ( c^{\dagger}_{1i} + c_{1i} )
		    \prod^{N_c}_{i=1} ( c^{\dagger}_{2i} - c_{2i} )
	= K Z_1 R_2
\label{p1122mp}
\end{equation}
Then one can show that
\begin{eqnarray}
 P^{\prime}_m \chi_{1i} P^{\prime}_m & = & P_m \chi_{1i} P_m =-\chi_{1i} ,  \nonumber   \\
 P^{\prime}_m \chi_{2i} P^{\prime}_m & = & P_m \chi_{2i} P_m = \chi_{2i}
\label{p1122mpact}
\end{eqnarray}
where, as usual,  $i=\infty $ is always excluded.
Then one can show $[P'_m,H_{12}]=0$ and $[P'_m,H_{1122}]=0$.
One can also show that  
$P^{\prime}_m Q_1 P^{\prime-1}_m=(N_c-1)-Q_1+2n_{1 \infty} $
and  $ P^{\prime}_m Q_2 P^{\prime-1}_m= N_c-Q_2 $,
 so $ P^{\prime}_m Q_t P^{\prime-1}_m= (2 N_c-1)-Q_t +  2 n_{1 \infty} $
 always map to the opposite total parity sector,
and it can only be used to establish the energy spectrum between opposite total parity sectors in the hybrid model Eq.\eqref{gr2hyv}.

Here, we summarize several salient features in Fig.\ref{color1122hy}.
Just from symmetry point of view, the hybrid model is always in GOE at a given total parity sector.
While the 3 GOEs in Table III is at a given parities sector $(Q_1,Q_2)$
which is conserved only at the $ q=4 $ SYK limit $ K/J=0 $.
As explained at Table III, the $ d_t=1+1 $ at $N$ odd is in the enlarged Hilbert space,
so the degeneracy can not be seen when one doing exact diagonalization in minimal Hilbert space.
Because this basis is the minimal original Hilbert space
without introducing $ \chi_{1 \infty} $ and $ \chi_{2 \infty} $ (see Appendix A).
When $N \pmod 4=1,3$, the GOE at $ K/J=0 $ is directly connected to the hybrid GOE,
this is why the GOEs in Fig.\ref{color1122hy}(a) and (c) are the two most robust ones against
the $H_{12}$ term among all the figures in Fig.\ref{color1122hy}.

In a sharp contrast, the GOE at $N \pmod 4=0$ can not be seen even at $K/J\rightarrow0$ limit.
Energy level with opposite parity are mixed,
so both parity sectors combine to behave like something slightly higher than ``Poisson''.
If one had done the exact diagonalization just in the total parity sector,
it may lead to the conclusion that
the $ q=4 $ 2-colored SYK satisfies Poisson hinting it might be integrable.
In reality, the quantum chaos is hiding inside the total parity
and need be dragged out by splitting it into the two separate parity sectors.
When comparing the knowledge in Sec. II B,
 one can find both $\langle \tilde{r}\rangle$ and $\langle \tilde{r}'\rangle$
 indeed match their prediction from mixed-2 GOE.
The ``fake'' Poisson will evolve to the GOE,
then a chaotic to non-chaotic transition from the GOE to the real Poisson.
As shown in Fig.\ref{color1122hy}(c),
the ``fake'' Poisson shows a nice plateau regime near $ q=4 $
whose length maybe used to quantitatively characterize
the stability of the quantum chaos near the $ q=4 $ side.

While the double degeneracy $d_t=2$ in the total parity sector at $N\pmod 4=2$
is in the minimal original Hilbert space, so can be seen in the exact diagonalization.
Any small $K$ breaks this degeneracy.
So the combination of $\langle\tilde{r}\rangle$
and the new universal ratio $\langle\tilde{r}'\rangle$
 first introduced in \cite{KAMSYK}  are needed to describe the evolution of the ELS.
 Especially,  $\langle\tilde{r}'\rangle$  is needed to quantitatively characterize the stability of the quantum chaos
 near the $ q=4 $ side.

\section{The 4-colored $q=4$ SYK}

Here, we take four colors  $ a=1,2,3,4 $ with $ q_1=q_2=q_3=q_4=1, N_1=N_2= N_3=N_4= N $,
and the 4-colored $q=4$ SYK can be written as
\begin{eqnarray}
    H_{1234} & = &  \sum^{N}_{i;j;k;l} J_{i;j;k;l} \chi_{1i} \chi_{2j} \chi_{3k} \chi_{4l}
\label{GR4}
\end{eqnarray}
where  $J_{i;j;k;l}$ are real and satisfy the Gaussian distribution with
mean value $ \langle J_{i;j;k;l} \rangle=0$
and variance $\langle J^2_{i;j;k;l} \rangle= 4 J^2/N^3 $.

\begin{figure}[!htb]
\centering
\includegraphics[width=0.9\linewidth]{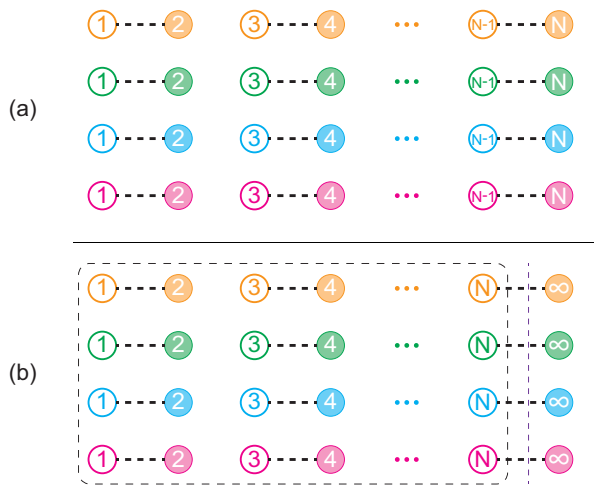}
\caption{ The 4-colored SYK with (a) $ N $ even and (b) $ N $ odd.
In (b), the long vertical dashed line separate the system from the four Majorana fermions
added at infinity. The  dashed box enclose the additional conserved quantity $ Q_{0t} $.
The Hilbert space is enlarged to $2^2=4$ times.
Simultaneously, there are also two more conserved parities. }
\label{fourcolor.eps}
\end{figure}


In contrast to the 2-colored cases,
the separate parity in each color $Q_a$, $a=1,2,3,4$ is not conserved any more,
but the parity of any sum of the two (there are 6 of them) are conserved.
Only 3 out of the 6 are independent.
Without losing any generality,
we can just pick the following 3: $ ( Q_{12}, Q_{23}, Q_{34} ) $ as a set.
Just like the 2-colored SYK model discussed in Sec. III,
an inter-color scheme description
introduce $ N $ complex fermions from the first two colors 1,2,
and another $ N $ complex fermions from the other two colors 3,4.
As to be shown in the Appendix B,
this construction using $ P_{12}$ and $P_{34} $ across the two of the four colors
is an alternative representation to discuss the symmetry class of the Hamiltonian.
In the following, we mainly focus on the intra-color scheme,
which keeps the conserved parity $(Q_{12},Q_{23},Q_{34})$ explicitly,
so there are 8 sectors which can still be regroup into
two sectors with two different total parities $Q_t=Q_{12}+Q_{34}$.
Both approached have their own advantages, and they are complementary to each other.

Due to a possible spectral mirror symmetry,
the 4-colored SYK models will be classified in 10-fold way.
The 10-fold way classification can be viewed as a generalization of Wigner-Dyson's 3-fold way,
and it is also known as Altland-Zirnbauer classification theory
\cite{Altland-Zirnbauer-1,Altland-Zirnbauer-2}.
Thanks to the one-to-one correspondence
between each ensemble and symmetric spaces in Cartan's classification,
we label ten RMT classes by its Cartan's name.
The classification need consider two anti-unitary operator $T_+$, $T_-$ and one unitary operator $\Lambda$.
$T_+$ commutes with the Hamiltonian as well as all compatible conserved quantities,
while $T_-$ and $\Lambda$ anti-commute with the Hamiltonian
but commute with all compatible conserved quantities.
Notice if both $T_+$ and $T_-$ exist, then there always exist $\Lambda=T_+T_-$,
but the converse is not true.
Ten RMT classes can be identified by following operator algebra:
i) $T_-$ and $\Lambda$ not exist,
if $T_+$ exist and $T_+^2=+1$, the Hamiltonian belongs class AI(GOE);
if $T_+$ exist and $T_+^2=-1$, the Hamiltonian belongs class AII(GSE);
if $T_+$ also not exist, the Hamiltonian belongs class AI(GUE).
In fact, this is reduced to Wigner-Dyson's 3-fold way discussed in Sec. III.
ii) either exist $T_+$, $T_-$ and $T_+^2=T_-^2$ or exist $\Lambda$,
if both $T_+^2=T_-^2=+1$, the Hamiltonian belongs class BDI (chGOE);
if both $T_+^2=T_-^2=+1$, the Hamiltonian belongs class CII (chGSE);
if only exist $\Lambda$,  the Hamiltonian belongs class AIII (chGUE).
As write in the parentheses, these are 3 chiral ensembles.
iii) either exist $T_+$, $T_-$ and $T_+^2\neq T_-^2$ or exist $T_-$,
if $T_+^2=-T_-^2=+1$, the Hamiltonian belongs class CI (BdG);
if $T_+^2=-T_-^2=-1$, the Hamiltonian belongs class DIII (BdG);
if $T_+$ not exist but $T_-$ exist with $T_-^2=+1$, the Hamiltonian belongs class D (BdG);
if $T_+$ not exist but $T_-$ exist with $T_-^2=-1$, the Hamiltonian belongs class C (BdG).
As write in the parentheses, these are 4 Bogoliubov-de Gennes ensembles.

\subsection{$ N $ even case}

In $N$ even case, just following the 2-colored SYK discussed above,
one can split the site $i$ into even and odd site (Fig.\ref{fourcolor.eps}a),
then introduce $ N_c=N/2 $ complex fermions for each color
$c_{1i}=(\chi_{1,2i}-i\chi_{1,2i-1})/\sqrt{2}$
and $c^{\dagger}_{1i}= ( \chi_{1,2i}+i \chi_{1,2i-1} )/\sqrt{2} $.
The particle-hole symmetry operator can be defined as
$P_1=K\prod^{N_c}_{i=1}(c^{\dagger}_{1i}+c_{1i})$
or $R_1= K \prod^{N_c}_{i=1}(c^{\dagger}_{1i}-c_{1i})$.
   It is easy to show $ P^2_1= (-1)^{\lfloor N_c/2\rfloor} $.
One can also show that
$P_1 c_{1i} P_1= \eta c^{\dagger}_{1i}$,
$P_1 c^{\dagger}_{1i} P_1= \eta c_{1i}$,
and $P_1 \chi_{1i} P_1= \eta \chi_{1i} $,
where $ \eta=(-1)^{\lfloor(N_c-1)/2\rfloor} $.
Neither the number operator of color 1 fermions
$ Q_1= \sum^{N_c}_{i=1} c^{\dagger}_{1i} c_{1i}$
nor its parity $ (-1)^{Q_1}$ is conserved.
Very similarly, one can construct $ P_2, P_3, P_4 $ and $R_2,R_3,R_4$.
So $ P_a, R_a $, $a=1,2,3,4$, can be used as the building blocks to construct all the possible operators.

\begin{figure}[!htb]
\centering
	\includegraphics[width=\linewidth]{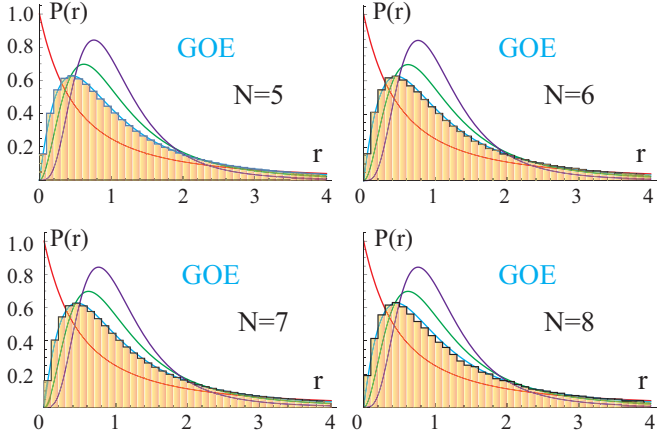}
	\caption{Distribution of the ratio of consecutive level spacings $P(r)$
	for 4-colored SYK model with various $N=5,6,7,8$.
	When $N\pmod 4=1,2,3,0$,
	all the bulk ELS show GOE which
	agrees with symmetry analysis summarized in Table IV.
	But they can be distinguished by different edge behaviours
	as shown in Fig.\ref{fourcoloredge}.
	The 4 background curves are $P(r)$
	of Poisson (red), GOE (blue), GUE (green), GSE (purple)	respectively.}
\label{fourcolorELS}
\end{figure}

Similar to the anti-unitary operator Eq.\eqref{p1122} introduced in 2-colored case,
we define:
\begin{align}
    P &=K\prod^{N_c}_{i=1}
	(c^{\dagger}_{1i}+c_{1i})(c^{\dagger}_{2i}+c_{2i})
	(c^{\dagger}_{3i}+c_{3i})(c^{\dagger}_{4i}+c_{4i})   \nonumber  \\
      &=KP_1P_2P_3P_4\>.
\label{p1234c}
\end{align}
It is easy to show that:
\begin{equation}
    P \chi_{ai} P
	=(-1)^{\lfloor\frac{N_c}{2}\rfloor + N_c} \eta \chi_{ai}
	= -\chi_{ai} ,\quad  a=1,2,3,4
\label{p1234cact}
\end{equation}
which leads to $[P,H_{1234}]=0$.
It is also easy to check that $P^2=1$ and $ P Q_a P^{-1}= N_c-Q_a$, $a=1,2,3,4$,
which automatically lead to
$P Q_{12} P^{-1}= 2N_c-Q_{12}$,
$P Q_{23} P^{-1}= 2N_c-Q_{23}$, and
$P Q_{34} P^{-1}= 2N_c-Q_{34}$.
Obviously, operator $P$ always maps $(Q_{12},Q_{23},Q_{34})$ to the same parity sector.

In fact, we can identify another anti-unitary operator:
\begin{align}
    P_m &= K \prod^{N_c}_{i=1}
	(c^{\dagger}_{1i}+c_{1i})(c^{\dagger}_{2i}+c_{2i})
	(c^{\dagger}_{3i}+c_{3i})(c^{\dagger}_{4i}-c_{4i})  \nonumber   \\
      &=  K P_1 P_2 P_3 R_4
\label{pm1234c}
\end{align}
which simply replace $P_4$ in Eq.\eqref{p1234c} by the $ R_4 $ operator.
It can be contrasted to the similar operator in the 2-colored case Eq.\eqref{p1122m}.
It is easy to show that:
\begin{eqnarray}
    P_m \chi_{ai} P_m
	& = & (-1)^{ \lfloor\frac{N_c}{2}\rfloor } \eta \chi_{ai}
	= (-1)^{N_c-1} \chi_{ai}, \quad a=1,2,3   \nonumber   \\
    P_m \chi_{4i} P_m
	& = & -(-1)^{ \lfloor\frac{N_c}{2}\rfloor } \eta \chi_{4i}
	=(-1)^{N_c} \chi_{4i}
\label{pm1234cact}
\end{eqnarray}
which indicates $\chi_{4i}$ has an opposite sign than the other 3 colors
and this opposite sign which leads to $ \{ P_m, H_{1234} \}=0 $.
It is also easy to check that $P^2_m=(-1)^{N_c}$ and
  $ P_m Q_a P^{-1}_m= N_c-Q_a$, $a=1,2,3,4 $
which automatically lead to
  $P_m Q_{12} P^{-1}_m= 2N_c-Q_{12}$,
  $P_m Q_{23} P^{-1}_m= 2N_c-Q_{23}$,
  $P_m Q_{34} P^{-1}_m= 2N_c-Q_{34}$.
Obviously, $ P_m $ also maps $ ( Q_{12}, Q_{23}, Q_{34} ) $ to the same parity sector.

\begin{figure}[!htb]
\centering
	\includegraphics[width=\linewidth]{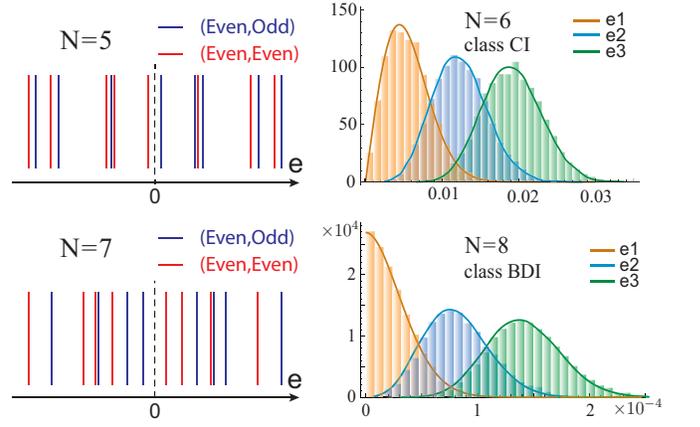}
	\caption{
	Distributions of the eigenvalues of 4-colored SYK model
	with a few smallest absolute values.
	For $N=6$ and $N=8$ case, each parity sector has mirror symmetry
	thus we calculate 3 smallest absolute values,
	and compare them with predications (solid lines) from RMT class BDI and CI
	and find RMT index $\alpha=0,1$ respectively.
	For $N=5$ and $N=7$ case, each parity sector has no mirror symmetry
	thus have no well defined $\alpha$ index.
	To show the absence of the mirror symmetry in the odd $ N $ case,
	we plot a few smallest absolute eigenvalues
	in the  $ ( Q_{12}, Q_{34} )=(+,+)$ or $(+,-)$ sector for a single random realization.	}
\label{fourcoloredge}
\end{figure}

  Now we find two anti-unitary operators, one commuting, another anti-commuting with $ H_{1234} $.
  From the two anti-unitary operators, one can define the chirality operator
  $ \Lambda= P P_m= P_4 R_4= (-1)^{Q_4} $
  which is an unitary operator anti-commuting with the Hamiltonian $ \{ \Lambda, H_{1234} \}=0 $.
  Of course, any $ (-1)^{Q_a}, a=1,2,3,4 $ work equally well as the unitary chirality operator.
  This is clearly intuitive, because $ H_{1234} $ in Eq.\eqref{GR4} contains one color each, so
  anti-commute with $ (-1)^{Q_a}, a=1,2,3,4 $.

  Overall, when combining $ P $ with $ P^2= 1 $ and $ P_{m} $ with $ P^2_{m}= (-1)^{N_c} $,
  one can see that when $N \pmod4=0,2$, $ N_c=N/2 $ is even or odd,
  so $ H_{1234} $ belongs to class BDI or class CI respectively.
  They have the RMT index $\beta=1$, $\alpha=0$ and $\beta=1$, $\alpha=1$ respectively.
  Both show the GOE bulk statistics,
  but with different edge exponent with $ \alpha=0 $ and $ \alpha=1 $ respectively.
  The fact $ P^2= 1 $ also lead to no level degeneracy in parity sector $(Q_{12}, Q_{23}, Q_{34})$.
  Since there is no operator can make connections between different parity sectors,
  the level degeneracy is $d_t=1$ in total parity sector $Q_t$.

\subsection{$ N $ odd case}


\subsubsection{In-complete classification with a missing conserved quantity }

When $N \pmod 4=1,3$, the above procedures for even $N$ needs to be modified.
In fact, one can still take the advantage of the above representation with $N$ even case
by adding decoupled Majorana fermions $ \chi_{a,N+1} =\chi_{a,\infty}$, $a=1,2,3,4$,
to make the parity conservations in $(Q_{12},Q_{23},Q_{34})$ explicitly (see Fig.\ref{fourcolor.eps}(b)).
Then one can still define $ P_a$, $a=1,2,3,4 $ and $ R_a$, $ a=1,2,3,4 $, $ P $ and $ P_m $
(therefore also the chirality operator $\Lambda_a=(-1)^{Q_a}$) with $N_c=(N+1)/2$ as before.
When $N \pmod 4=3$, $ N_c $ is even, $ P^2=1, P^2_m=1 $, it is in class BDI;
when $N \pmod 4=1$, $ N_c $ is odd, $ P^2=1, P^2_m=-1 $, it is in class CI.
Unfortunately, the conclusion is {\sl in-correct}.
This could be expected that the Hilbert space is enlarged $4$ times.
Simultaneously, there should also two more conserved parities
when comparing with the total number of conserved quantities in inter-color scheme which is 2.
We only have 3 as $ ( Q_{12}, Q_{23}, Q_{34} ) $,
so one conserved quantity is still missing,
and we will find this missing parity in Sec. V B2.

When $N$ is odd, one may also use the following operator:
\begin{equation}
    P_z = K Z_1 P_2 Z_3 P_4
\label{pm1234czz}
\end{equation}
which simply replace $ P_1, P_3 $ in Eq.\eqref{p1234c} by $ Z_1, Z_2 $ operator.
So it will play a complementary role as $P$, which will be analyzed in the following.
One can show that:
\begin{equation}
 P_z \chi_{ai} P_z  = - (-1)^{ \lfloor\frac{N_c}{2}\rfloor + N_c } \eta \chi_{ai}= \chi_{ai},
	\quad a=1,2,3,4
\label{pm1234czzact}
\end{equation}
where, as usual, $ i=\infty $ is always excluded.
  It is also easy to check that $ [ P_z, H_{1234}]=0, P^2_z=-1  $ and
  $ P_z Q_a P^{-1}_z= N_c-1-Q_a + 2 n_{a \infty}, a=1, 3 $
  and  $ P_z Q_a P^{-1}_z= N_c-Q_a, a=2, 4 $
  which automatically lead to
  $ P_z Q_{12} P^{-1}_z= 2N_c-1-Q_{12}+ 2 n_{1 \infty}$,
  $ P_z Q_{23} P^{-1}_z= 2N_c-1-Q_{23}+ 2 n_{3 \infty}$,
  $ P_z Q_{34} P^{-1}_z= 2N_c-1-Q_{34}+ 2 n_{3 \infty}$.
  Obviously, operator $ P_z $ maps  $ ( Q_{12}, Q_{23}, Q_{34} ) $ to
  a different parity sector $ ( Q_{12}+1, Q_{23}+1, Q_{34}+1 ) $.
  However, both sets have the same total parity $ Q_t= Q_{12} + Q_{34} $.

\subsubsection{Complete classification by finding the missing conserved quantity $ Q_{0t} $.}

Unfortunately, the above classification disagrees with our exact diagonalization results, especially on edge exponents.
It is important to resolve the discrepancy.
It turns out that it missed the additional conserved quantity $Q_{0t}$,
which is the parity in the square box in Fig.\ref{fourcolor.eps}(b).
In the 2-colored cases discussed in Sec. III and IV,
it is also a conserved quantity, but it does not commute with parity $ Q_1 $ and parity $ Q_2 $,
so can not be used in the complete set of the conserved quantities.
Here, it commutes with parities $ Q_{12}, Q_{23}, Q_{34} $.
So $Q_{0t}$ is the (so far missing) additional member of the complete set of the conserved quantities
  $ ( Q_{12}, Q_{23}, Q_{34}, Q_{0t} ) $.
  From Fig.\ref{fourcolor.eps}(b), it is easy to see
\begin{equation}
  Q_t= Q_{12}+Q_{34}=Q_{0t}+ n_{12 \infty}+ n_{34 \infty}
\end{equation}
where $ n_{12 \infty}$, $n_{34 \infty}$ may not be able to be
conveniently expressed in terms of complex fermions $c_i, i=1,2,3,4 $,
but can always be concisely expressed in terms of Majorana fermions
$ n_{12 \infty}= \frac{1}{2}- i \chi_{1 \infty} \chi_{2 \infty}$,
$ n_{34 \infty}= \frac{1}{2}- i \chi_{3 \infty} \chi_{4 \infty} $.
Then one can show
\begin{eqnarray}
  P Q_{0t} P^{-1} & =  &4 N_c+2- Q_{0t}-2 (   n_{12 \infty}+ n_{34 \infty} )    \nonumber  \\
  P_m Q_{0t} P^{-1}_m & =  &  4 N_c+ 1- Q_{0t}-2 n_{12 \infty}
\end{eqnarray}
So one can see that $ P_m $ operator {\sl changes} the parity of $ Q_{0t} $.
This fact eliminates $ P_m $ as the valid operator and leaves $ P $ as the only valid one,
so there is no spectral mirror symmetry in given parities sector $(Q_{12},Q_{23},Q_{34},Q_{0t})$.
Because of $P^2=1$ and lacking of spectral mirror symmetry,
the ELS is class AI (GOE), and no edge exponent can be defined for GOE.

One can also show
\begin{equation}
  P_z Q_{0t} P^{-1}_z
	\!=\!  4 N_c-Q_{0t}+2(n_{1 \infty}\!+n_{3 \infty}\!-n_{12 \infty}\!-n_{34 \infty})\>,
\end{equation}
which shows that $ P_z $ conserves $(Q_t,Q_{0t})$.
This fact shows that in a given total parity sector $(Q_t,Q_{0t})$,
the energy level has two fold degeneracy $ d_t=2 $.
This result could be useful when a quadratic term like Eq.\eqref{gr4all},
which breaks the parities, but still keeps the total parity.

In summary,
when $N \pmod 4=0$, it is in class BDI;
when $N \pmod 4=2$, it is in class CI.
In additional to the bulk, they also have the edge exponents.
When $N \pmod 2=1,3$, it is in class AI(GOE) with $ d_t=2 $, and no edge exponent can be defined.
These theoretical results are confirmed by the exact diagonalization shown
in Fig.\ref{fourcolorELS} for the bulk and Fig.\ref{fourcoloredge} for the hard edge behaviour.


\renewcommand{\arraystretch}{1.5}
\renewcommand\tabcolsep{6.1pt}
\begin{table}
\label{tab:4c}	
\caption{The ELS and degeneracy of the four colored SYK model.
The degeneracy $ d=1 $ is a given parity sector $ ( Q_{12}, Q_{23}, Q_{34}, Q_{0t} ) $.
$ Q_{0t} $ is defined only when $ N $ is odd.
The total degeneracy $ d_t $ is at a total parity sector $  ( Q_t, Q_{0t} ) $.
When N is odd, $ P_z $ operator in Eq.\ref{pm1234czz} maps $ ( Q_{12}, Q_{23}, Q_{34} ) $ to
a different parity sector $ ( Q_{12}+1, Q_{23}+1, Q_{34}+1 ) $.
However, both sets have the same total parity $ ( Q_t, Q_{0t} ) $. So $ d_t=2 $.
When doing exact diagonalization $ P_{12} $ and $ P_{34} $  basis, both sets $ d_t=2 $ can be seen and were shown
in Fig.\ref{color1234hybrid}a,c (see also Appendix B). }
\centering
\begin{tabular}{  c | c | c  | c | c  }
\toprule
  $N \pmod 4$	    & 0	    &   1   &   2   &  3  \\  \hline
  ELS		    & BDI   &  AI   &  CI   & AI \\  \hline
  $(\beta,\alpha)$  & (1,0) & (1,-) & (1,1) & (1,-)      \\  \hline
  $ (Q_{12},Q_{23},Q_{34}, Q_{0t} ) $
     & $d=1$     & $d=1$	    & $d=1$	&  $d=1$       \\  \hline
  $ ( Q_t, Q_{0t} ) $    &  $d_t=1$   & $d_t=2$   & $d_t=1$	&  $d_t=2 $   \\   \hline
\end{tabular}
\end{table}


\section{The hybrid 4-colored $ q=2 $ and $ q=4 $ SYK model}

Similar to the 2-colored cases, in the following,
we will discuss the parity $(Q_{12},Q_{23},Q_{34},Q_{0t})$
violating hybrid 4-colored SYK model Eq.\ref{gr4all}.
It still conserves the total parity $(Q_t,Q_{0t})$.
It can  be used to study the stability of quantum chaos
and Kolmogorov-Arnold-Moser theorem in the $ f=4 $ colored SYK \cite{KAMSYK}.
Furthermore, one can demonstrate the importance of identify the maximal symmetry,
the largest conserved quantities and the smallest Hilbert space
to do the correct classifications in the RMT.
A small perturbation $ K/J \rightarrow 0 $ limit which breaks
$ ( Q_{12},Q_{23}, Q_{34} ) $, but conserves $ Q_t=Q_{12} + Q_{34} $ may also be used to drag out the
rich and novel physics encoded in the Table IV.
This kind of small perturbation may also be used to probe the interior
of a dual black hole in the bulk \cite{pure}.

\begin{figure}[!htb]
\centering
\includegraphics[width=\linewidth]{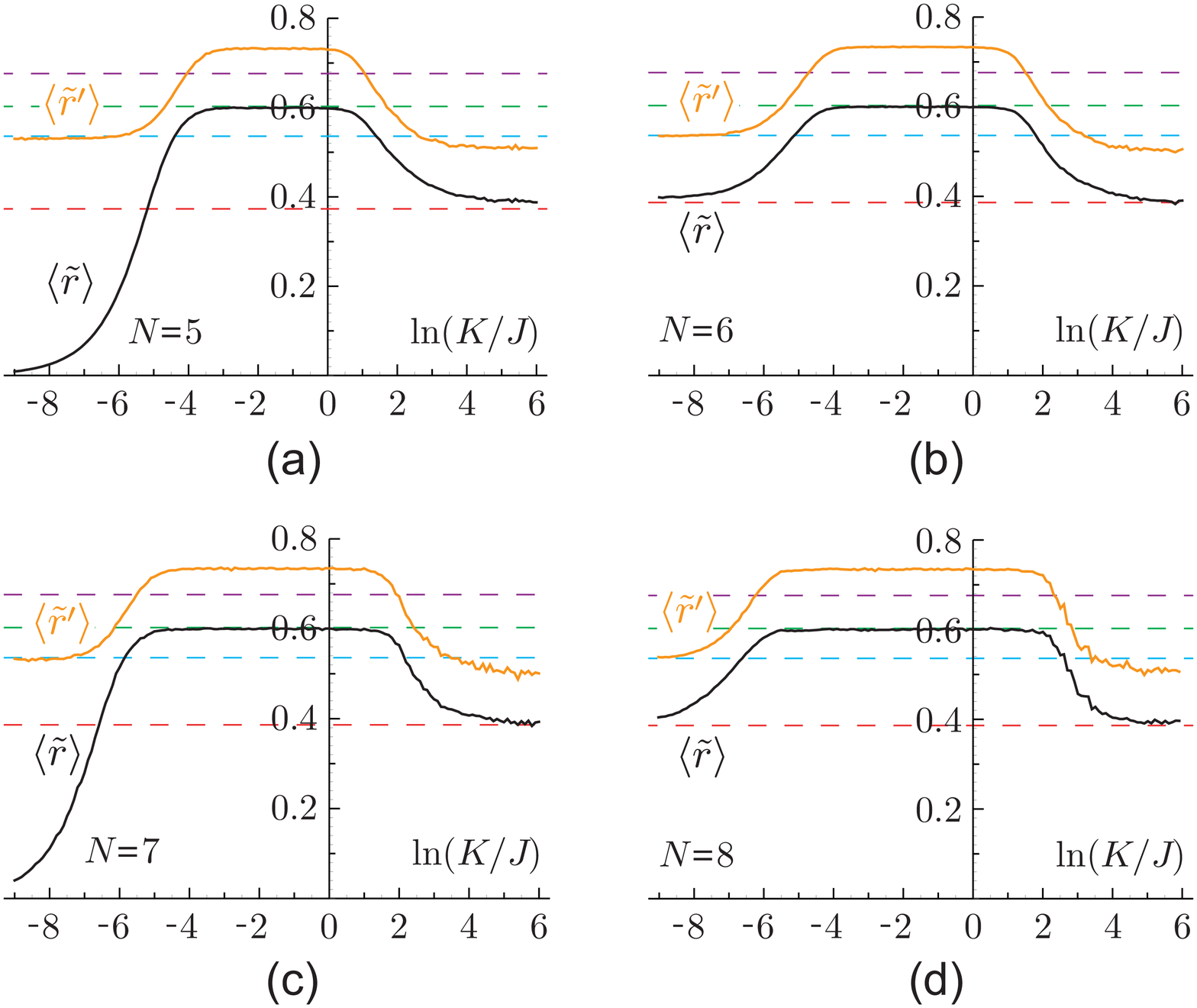}
\caption{(Color online)
The mean value of the $\tilde{r}$-parameter and $\tilde{r}'$-parameter for
the 4-colored hybrid SYK models with $N=5,6,7,8$.
All data are taken at a given total parity sector $Q_t=Q_1+Q_2$,
and averaged over $400$, $200$, $10$, $6$ samples respectively.
For $ N \pmod 4=1,3$ in (a) and (c), the $ H_{1234} $ (namely, at $ K=0 $)
has two-fold degeneracy $ d_t=2 $ in the total parity sector.
$\langle\tilde{r}\rangle$ (black curve) for NN ELS is rapidly changing,
but $\langle\tilde{r}'\rangle$ (orange curve) for NNN ELS shows a nice GOE plateau near $K=0$.
The hybrid SYK is in GUE in some range near $ K/J =1 $
and there is a chaotic to non-chaotic transition from the GUE to the Poisson as $ K/J $ increases.
As shown in \cite{KAMSYK}, when $\langle\tilde{r}\rangle$ is at the GUE ($\beta=2$) value $0.6027$,
$\langle\tilde{r}'\rangle$ would be close to the value $0.7344$.
When $\langle\tilde{r}\rangle$ is in Poisson value $0.3863$,
the  $\langle\tilde{r}'\rangle$ would be close to $1/2$
which is slightly  below the $\langle\tilde{r}\rangle$'s GOE value $0.5359$.
For $ N \pmod 4=2,0 $ in (b) and (d), as explained in Sec.IIB and the text,
the very slightly above the Poisson values on the left near the $ q=4 $ side
is due to the mixing of 4 un-correlated parity sectors.
The quantum chaos in the GOE is hidden in this ``fake'' Poisson
and can be dragged out by doing ELS on a given parity sector $(Q_{12},Q_{23},Q_{34})$.
While, the Poisson one on the right near $q=2$ side is a true one.
As listed in the Table II, the NNN ELS of the 4 mixed un-correlated sectors
lead to $\langle \tilde{r}'\rangle\approx0.535$ which is quite close to
the GOE value of $\langle\tilde{r}\rangle=0.5359 $.
Indeed, $\tilde{r}'$ shows a plateau on GOE at small $ K/J $.
}
\label{color1234hybrid}
\end{figure}

A hybrid 4-colored $ q=2 $ and $ q=4 $ SYK model is:
\begin{align}
    H^{Hb}_{1234}
	\!=\!\!\sum^{N}_{i,j,k,l}\! J_{i;j;k;l} \chi_{1i} \chi_{2j} \chi_{3k} \chi_{4l}
	+ i \!\! \sum^{N}_{i,j;a<b} \!\!\! K_{i;j} \chi_{ai} \chi_{bj}
\label{gr4all}
\end{align}
Of course, other hybrid 4-colored models can also be constructed,
but Eq.\eqref{gr4all} is the most democratic one among all the four colors.
In fact, shown in Fig.\ref{color1234hybrid} is our exact diagonalization in a slightly generalized model
$H^{Hb\prime}_{1234}= \sum_{i,j,k,l} J_{ijkl} \chi_{1i} \chi_{2j} \chi_{3k} \chi_{4l}
  +i\sum_{i,j;a<b} K^{ab}_{ij} \chi_{ai} \chi_{bj}$
where $ K^{ab}_{ij} $ also depends on colors, but satisfy the same distribution.
Our exact diagonalization on Eq.\eqref{gr4all} leads to similar results
with slightly more noises on a given distribution.

Now we can apply the particle-hole transformation $P$ and $P_m$ to it.
Although $(Q_{12},Q_{23},Q_{34})$ are not conserved anymore,
the total parity $(-1)^{Q_t}$ (and $ (-1)^{Q_{0t}} $ when $ N $ is odd) remains conserved.
It is also easy to see that $  \{ P_{m}, H_{12} \} =  \{ P_{m}, H_{13} \}=  \{ P_{m}, H_{23} \} = 0 $
and $ [P_{m}, H_{14} ]=[P_{m}, H_{24} ]=[P_{m}, H_{34} ]=0 $.
So $ P_{m} $ neither commutes nor anti-commutes with $H^{Hb}_{1234}$.
Because $[P,H_{1234}]=[P_z,H_{1234}]=0$  and $\{P,\sum_{ab}H_{ab}\}=\{P_z,\sum_{ab}H_{ab}\}=0$,
the hybrid 4-colored SYK does not have any symmetry anymore.
This is in sharp contrast to the hybrid 2-colored SYK Eq.\ref{gr2hyv}
where one can still identify a conserved quantity $P_m$ Eq.\ref{p1122m}.
Just from symmetry point of view, the hybrid 4-colored SYK belongs to the class A (GUE),
so the ELS may satisfy GUE for any ratio of $ K/J $.
When performing the exact diagonalization, we need to look at a given total parity $  (-1)^{Q_t} $.
However, the Kolmogorov-Arnold-Moser theorem shows that as $ K/J $ increases to $ (K/J)_c $,
there maybe a chaotic to non-chaotic transition from the GUE to the Poisson.
Our exact diagonalization studies shown in Fig.\ref{color1234hybrid} confirms this picture.

\subsection{$ N $ even case}

It is instructive to look at $K/J\to0$ limit in Fig.\ref{color1234hybrid}(b) and (d).
If one just focus on the total parity sector $(-1)^{Q_t}$,
As shown in Table IV, there is a no level degeneracy $ d_t=1 $.
When $N \pmod 4=0,2$, the exact diagonalization is performed in total parity sector $Q_t$.
There are four separate parities in $(Q_{12},Q_{23},Q_{34})$
falling in the same total parity sector $Q_t$,
and these 4 sectors are completely independent of each other.
Since no level repulsion among the 4 sets of energy levels,
then the ELS may start to show something similar to (in fact, very slightly above) Poisson \cite{also}.
This is indeed the case shown in Fig.\ref{color1234hybrid}(b) and (d).
Naively, it could mislead to the conclusion that $ H_{1234} $ maybe integrable when $N \pmod 4=0,2$.
Note that here it is 4 sectors are mixed together in the same total parity sector $ Q_t $,
while only two sectors are mixed in the 2-colored case,
so the 4-colored case is expected to be more close to the Poisson than the 2-colored case.
Indeed, in Sec. II B and Table II, we show that mixed 4 UCID GOE, GUE, or GSE,
all lead to $\langle \tilde{r}\rangle\approx0.39$ and $\langle \tilde{r}'\rangle\approx 0.535$,
which agree with the Fig.\ref{color1234hybrid}(b),(d).
So the values of $\langle \tilde{r}\rangle$ and $\langle \tilde{r}'\rangle$  in Fig.\ref{color1234hybrid}(b),(d)
at small $K/J$ mean 4 un-correlated GOE are mixed together.
However, as shown in Fig.\ref{fourcolorELS}, if one do the ELS on separate parities $(Q_{12},Q_{23},Q_{34})$, then the ELS shows its real face: class BDI and class CI.

As $ K/J $ increases to $ (K/J)_c $, there is a crossover from the ``fake'' Poisson to GUE,
then followed by a chaotic to non-chaotic transition from GUE to the real Poisson near $ q=2 $.
As shown in Fig.\ref{color1234hybrid}(b) and (d),
the ``fake'' Poisson shows a nice plateau regime in both values of $\langle \tilde{r}\rangle$ and $\langle \tilde{r}'\rangle$ near $ q=4 $
whose length maybe used to quantitatively characterize
the stability of the quantum chaos near the $ q=4 $ side.

\subsection{$ N $ odd case}

Similarly, one can first look at $K/J\to0$ limit in Fig.\ref{color1234hybrid}(a) and (c).
As shown in Table IV,
when $N \pmod 4=1, 3$,
due to the existence of the $P_z$ operator which maps
$( Q_{12},Q_{23}, Q_{34}, Q_{0t}  ) $ to $( Q_{12}+1,Q_{23}+1, Q_{34}+1, Q_{0t}  ) $,
there is a double degeneracy $ d_t=2 $.
The degeneracy is broken by any non-zero $ K/J$.
However, as shown in the last section,
when doing the ELS on separate parities $(Q_{12},Q_{23},Q_{34},Q_{0t})$,
the ELS shows GOE (Fig.\ref{fourcolorELS}b,d).
The evolution and fine structures characterized by
NN ratio $\langle\tilde{r}\rangle$ and NNN ratio $\langle\tilde{r}'\rangle$
are shown in Fig.\ref{color1234hybrid} a,c.
Especially, $\langle\tilde{r}'\rangle$ is needed to quantitatively
characterize the stability of the quantum chaos near the $ q=4 $ side.
As $ K/J $ increases to $ (K/J)_c $, there is a crossover from the GOE to GUE,
then followed by a chaotic to non-chaotic transition from GUE to the real Poisson near $ q=2 $.

\section{Perspectives and Discussions}

As mentioned in the introduction, during the last decade,
since the discovery of the topological insulators \cite{kane,zhang},
there are also extensive research activities on the classifications of
topological phase of matter which break no symmetries \cite{tenfold,wen}.
These phases also split into two classes:
interacting symmetry protected topological (SPT) phases
with trivial bulk order (short-range entanglement)
and symmetry enriched topological (SET) phases
with non-trivial  bulk topological order (long-range entanglement) \cite{tenfold,wen}.
In some special cases, the Hamiltonian whose exact ground states
show such SPT or SET orders can be constructed, but these Hamiltonian,
in general, involves highly non-local interactions
which are needed to stabilize such states.
In most cases, the Hamiltonians which may host these phases are not known,
the classifications are purely symmetry based.
For a general simple experimental accessible Hamiltonian,
these states may have much higher energy than conventional symmetry broken states.

A dual vortex method (DVM) was developed in \cite{pq1,frus1,frus2,frus3}
to classify all the possible Mott insulating phases
of interacting bosons hopping in a various 2d lattices
at generic commensurate filling factors  $ f=p/q $
($ p, q $ are relative prime numbers).
The DVM is a magnetic space group (MSG) symmetry-based approach
which, in principle, can be used to classify all the possible phases
and phase transitions in a extended Boson Hubbard (EBHM) model.
But if a particular phase identified by the DVM will become a stable ground state
or not depends on the specific values of all the possible parameters in the EBHM.
This kind of question can only be addressed by a microscopic approach
such as Quantum Monte-Carlo simulations on a specific Hamiltonian.
The combination of both methods are needed to completely understand quantum phases
and phase transitions in the EBHM.
Similar kind of approach was extended to 3d (called vortex condensation approach)
to classify SPT phases in 3+1 D  \cite{senthil,tenfold,wen}.

The possible organization patterns of matter can also be classified
from a different perspective: they can also be classified
by how quantum information are scrambled in the system.
So in this paper, we took a different route and achieved different goals:
we classify different types of quantum chaos
and quantum information scramblings in the colored SYK models
instead of its topological equivalent classes by using the RMT.
Here, we already wrote down a realistic colored SYK Hamiltonian,
and identify its maximal symmetries, the largest number of conserved quantities
(which are various fermion parities) and the smallest (irreducible)  Hilbert space.
Especially, one must also exhaust all the possible anti-unitary operators
which commutes or anti-commutes the Hamiltonian.
There are also two kinds of such anti-unitary operators,
the first kind keep all the conserved quantities in the same sector,
 the second kind map out of the sector.
 The former leads to the RMT classification, the latter establishes the connections between different sectors, therefore the degeneracy of the energy levels.
 If any symmetry or conserved quantity or any operator is missed,
 it can lead to  misleading results in both classifications and exact diagonalization results.
 We achieved such a goal in classifying the quantum chaos in colored SYK with 2 and 4 colors and balanced number of Majorana fermions among different colors.
 The color degree of freedoms may also be promoted to a global symmetry $ G $,
 then the parities are promoted to various conserved quantities,
 so it is also interesting to see how the color degree of freedoms compared to the SYK model with a global $ G=O(M) $ or $ G=U(M) $ symmetries \cite{tensorglobal}.
 As shown in \cite{graph}, there are some still unknown relations between colored SYK and the colored (Gurau-Witten) tensor model.
 The method can also be applied to do the RMT classifications of colored tensor models \cite{tensorrev}.

 It is interesting to note that the RMT was originally proposed to study the many body energy level correlations of a nuclei with a large atomic number to hold large number of electrons \cite{WD1,WD2}. Then it was also used to classify the quantum chaos of non-interacting
 electrons moving in a random potential which may show metal to Anderson insulator (MIT) transition \cite{efetov}.
 There is a corresponding chaotic to non-chaotic transition where the single particle
 ELS satisfies Wigner-Dyson in the metal, while Poisson in the Anderson insulator.
 RMT was first applied to QCD in \cite{QCD-a} and was classified in \cite{QCD-b}.
 In the presence of pairing such as colored superconductivity in QCD,
 fermion numbers are no longer conserved, only the fermion parities are conserved.
 So our method may also be applied to do the RMT descriptions of colored superconductivity in QCD
 \cite{QCD-1,QCD-2}.
 On the other hand, the topological equivalent classes of the SPT phases of non-interacting electrons such as topological insulators or superconductors can be classified with the same symbols as the 10-fold way of the RMT \cite{kane,zhang} in terms of the two possible anti-unitary operators.  However, for  many body interacting electron systems, the anti-unitary operators are not enough,
 the SPT or SET phases maybe classified by using more
 advanced mathematical tools such as co-homology, corbordism and tensor categories \cite{tenfold,wen} which were already used
 in rational conformal field theory (RCFT) and also topological quantum computing.

 As shown in this paper, the color degree of freedoms make dramatic differences.
 This is due to the color degree of freedoms leads to more conserved parities and also more anti-unitary or unitary operators.
 As shown in the text, even or odd number of Majorana fermions seem make a lot of differences.
 This is also due to  even or odd lead to different number of mutually conserved parities and also different anti-unitary or unitary operator
 contents. In retrospect, the multi-channel Kondo models lead to dramatic differences than a single channel
 Kondo models \cite{stevekondo,Malkondo,kondoye12345,kondoye123452} where the channel index plays a similar role as the color index here.
 Furthermore,  only the boundary conditions changing in odd number of fermions
 lead to non-Fermi liquid behaviours, therefore absence of any quasi-particles.
 We expect it leads to chaotic behaviours.
 While the boundary conditions changing in even number of fermions
 lead to Fermi liquid behaviours with well defined quasi-particle excitations.
 We expect it leads to non-chaotic behaviours.
 Of course, the odd number of  Majorana fermions lead to non-trivial topology and
 play dramatic roles in the classifications of topological phases of matter \cite{tenfold,wen},
 while usually, even number of Majorana fermions do not.






  As presented in the introduction, there are at least two different ways to characterize the quantum chaos or quantum information scramblings.
  One way is to use the Lyapunov exponent (or spectrum) to characterize the quantum information scramblings, it can be extracted through
  evaluating OTOC at an early time $ t_d < t < t_d \log N=t_s$
(namely, between dissipation time and Ehrenfest time) by a large $ N $ expansion.
  It is insensitive to $N \pmod 8$ Bott periodicity and also the ground state degeneracy.
  Another way is to use RMT to  characterize energy level statistics or spectral form
  factor in a 10-fold way at a finite and large enough $ N $ which shows $N \pmod 8$ Bott periodicity and also the ground state degeneracy.
  The many body energy level spacing $ \Delta E \sim e^{-N} $, so the RMT describes the
  energy level correlations at the Heisenberg time scale $ t_{H} \sim 1/\Delta E \sim e^{N} $.
  Because the wide separation of the two time scales $ t_s $ and $ t_H $,
  it remains an open problem to explore the relations between the two schemes.
  It was believed that the two schemes are complementary to each other to characterize quantum chaos of a system from
  different perspectives. So it remains an outstanding problem to investigate the connections between
  the results achieved by RMT here with those achieved by the OTOC in \cite{Gross}.

 The colored SYK models may also be experimentally realized in various cold atom \cite{coldwire},
 cavity QED systems \cite{KAM} or solid state system \cite{solid-1,solid-2,solid-3}
 where there maybe always color degrees of freedom.
 They naturally stands for with different band indices in a material.
 This could  achieve the lofty goal of investigating
 various exotic properties of quantum black holes just in a conventional lab on earth.

 Note added: Recently, two of the authors found the 2 colored SYK models may find its application in transverse 
 wormholes with an attractive interaction \cite{kamsykcolorim}.

\section*{Acknowledgements }

J. Y thank C. Xu and S. Shenker for helpful discussions.
F.S and J.Y acknowledge AFOSR FA9550-16-1-0412 for supports.
This research at KITP was supported in part by the National Science Foundation under Grant No. NSF PHY-1748958.
Y.Y and W.L were supported by the NKBRSFC under grants Nos. 2011CB921502, 2012CB821305, NSFC under grants Nos. 61227902, 61378017, 11311120053.

\appendix
\section*{Appendix}

  In the three appendices, we will give an alternative inter-color pairing presentation to classify
  the quantum chaos in the 2-colored (Fig.\ref{twocolorapp.eps}), 4-colored cases (Fig.\ref{fourcolorapp.eps})
  and also the corresponding hybrid colored SYK models respectively.
  It maybe a quite natural approach at the first sight.
  It may also be the most convenient and economic basis to do exact diagonalization,
  because by pairing across different colors,
  one can construct the minimal Hilbert space to do the exact diagonalization no matter $ N $ is even or odd.

  However, for $N$ is even,
  due to the hidden of the separate parity conservations
  in $ ( Q_1, Q_2) $ in the two color case and $ ( Q_{12},Q_{23}, Q_{34} ) $ in the 4 color case,
  special cares are needed to identify the complete set of conserved quantities,
  the relevant operators to perform the classification and  derive the degeneracy.

  For $ N $ is odd, one may do the classification in the minimum Hilbert space
  with a given $ Q_{12} $ in the 2-colored case
  or with a given $ ( Q_{12}, Q_{34} ) $ in the 4-colored case.
  To be compared to the results achieved in the main text with the intra-color representation,
  by adding Majorana fermions at $ \infty $, one may also do the classification
  in the 2-times enlarged Hilbert space $ ( Q_{12}, \tilde{Q}_{12} ) $ in the 2-colored case
  and in the 4-times enlarged Hilbert space $ ( Q_{12}, \tilde{Q}_{12},  Q_{34}, \tilde{Q}_{34} ) $ in the 4-colored case.

  Obviously, the inter-color scheme is specialized to the balanced case only,
  can not be generalized to the imbalanced case.
  While the approach used in the main text can be easily generalized to the imbalanced case.
  It is constructive to compare the two (when $ N $ is even) or  three (when $ N $ is odd) different classification schemes 
  which not only lead to the same conclusions,
  but also bring additional considerable insights into the physical picture which may have broad
  impacts to other problems.

When $ N $ is even, the two schemes are:
(a) for the 2-colored case, $(Q_1,Q_2)$ intra-color scheme in the main text
and the $(Q_1,Q_2)$ inter-color scheme in the appendix.
(b) for the 4-colored case, $(Q_{12},Q_{23},Q_{34})$ intra-color scheme in the main text
and the $(Q_{12},Q_{23},Q_{34})$ inter-color scheme in the appendix.

When $ N $ is odd, the three schemes are:
(a) for the 2-colored case, $(\tilde{Q}_1,\tilde{Q}_2)$ intra-color scheme in the main text,
the $Q_{12}$ minimum Hilbert inter-color space in the appendix,
and $(Q_{12},\tilde{Q}_{12})$ inter-color scheme in the appendix.
(b) for the four color case,
$(\tilde{Q}_{12},\tilde{Q}_{23},\tilde{Q}_{34},Q_t)$ intra-color scheme in the main text,
the $(Q_{12},Q_{34})$ minimum Hilbert inter-color space in the appendix
and $(Q_{12},\tilde{Q}_{12},Q_{34},\tilde{Q}_{34})$ inter-color scheme in the appendix.


\subsection*{A. 2-colored SYK model: 3-fold way and operator $ P_{12} $ across the two colors}

At first sight, for both $ N $ even or odd,
one can always introduce $N$ complex fermions by combining the two flavors
$c_i=(\chi_{1i}-i \chi_{2i} )/\sqrt{2}$,
$c^{\dagger}_i= ( \chi_{1i}+i \chi_{2i} )/\sqrt{2} $
and define the particle-hole symmetry operator
to be $ P_{12}
= K \prod^{N}_{i=1} ( c^{\dagger}_i + c_i )
=K \chi_{1,1} \chi_{1,2} \cdots \chi_{1, N}  $
only involving the color 1
In fact, $ R_{12}=P_{12} (-1)^{Q_{12}}
= K \prod^{N}_{i=1} ( c^{\dagger}_i -c_i )
= K i \chi_{2,1} i \chi_{2,2} \cdots i\chi_{2, N}  $
involving only the color 2 works equally well, but it will not lead to new symmetry.
It is easy to show $ P^2_{12}= (-1)^{\lfloor N/2\rfloor} $.
One can also show that
$  P_{12} c_i P_{12}= \eta c^{\dagger}_i$,
$P_{12} c^{\dagger}_i P_{12}= \eta c_i$,
$P_{12} \chi_{ai} P_{12}= \eta \chi_{ai}$, $a=1,2 $
where $ \eta=(-1)^{\lfloor(N-1)/2\rfloor} $.
The total number of fermions $ Q_t= \sum^{N}_{i=1} c^{\dagger}_i  c_i $
is not a conserved quantity, but the parity $ (-1)^{Q_t} $ is in $ H_{1234}$.
Then  $P_{12} Q_t P^{-1}_{12}= N-Q_t $ which justifies $ P_{12} $ as an anti-unitary particle-hole transformation. And $ P_{12} $ also commutes with the Hamiltonian $ [ P_{12}, H_{1234}] =0 $.
It seems indicate the ELS is the same as the complex fermion SYK case discussed previously
\cite{MBLSPT,CSYKnum,KAMSYK}:
(1) $N \pmod 4=1,3$ it is GUE with $ d_t=1 $ in a given parity $ Q_t $,
(2) $N \pmod 4=0$, it is GOE with $ d_t=1 $ in a given parity $ Q_t $
(3) $N \pmod 4=2$, it is GSE with $ d_t=2 $ in a given parity $ Q_t $.
Unfortunately, these results are in-consistent with those listed in Table III.
In the following, we study how to remedy the problems.

\begin{figure}[!htb]
\centering
\includegraphics[width=0.9\linewidth]{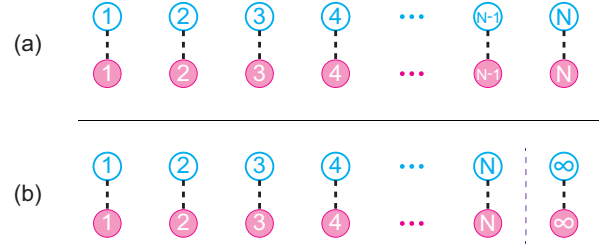}
\caption{ The two colors with inter-color pairings  (a) $ N $ even (b) $ N $ odd.
  . In contrast to the intra-color pairings in the main text.
  Color 1 are all real, while color 2 are all imaginary.
  In (b), the long vertical dashed line
  separate the system from the two Majorana fermions added at infinity.
  One doubles the Hilbert space in (b). Simultaneously, there is also one more conserved parity.
  Compare with Fig.\ref{twocolor.eps}. }
\label{twocolorapp.eps}
\end{figure}

{\sl (a) N is even }

Obviously, the total parity $ Q_{12}=Q_1 + Q_2 $ is not enough,
because  $ ( Q_1,Q_2 ) $ are separately conserved.
It may not be convenient to express $ ( Q_1,Q_2 ) $  in terms of the complex fermions
   $ c_i, c^{\dagger}_i $ in this basis, but can be most conveniently expressed in terms of the Majorana fermions
   as shown below.

   To fix this problem, we may still take $ Q_1=\sum^{N_c}_{i=1} c^{\dagger}_{i1} c_{i1}=
   \sum^{N_c}_{i=1} ( \frac{1}{2}- i \chi_{1,2i} \chi_{1,2i-1}  ) $ where $ N_c=N/2 $ as defined in Sec.2(a).
   Note that in the inter-color scheme Fig.\ref{twocolorapp.eps}a, $ Q_1 $ becomes complex,
   but remains Hermitian, so its eigenvalue remains positive.
   This is the price one must pay in this cross-color representation which make the Hamiltonian real,
   (namely, $ [K,H]=0 $), but  many other operators such as the $ Q_1 $ are complex.
   Similar results hold for $ Q_2 $.
   Then one can show  $ P_{12} Q_1 P^{-1}_{12}= N_c-Q_1$,  $P_{12} Q_1 P^{-1}_{12}= N_c-Q_2 $
   which recovers $ P_{12} Q_t P^{-1}_{12}= N-Q_t $.
   When $ N \pmod 4=0 $,  $ P_{12} $ maps $(Q_1,Q_2) $ to the same sector, $ P^2_{12}=1 $, it is in GOE.
   When $ N \pmod 4=2 $, $ P_{12} $ maps $(Q_1,Q_2)$ to  $(Q_1+1,Q_2+1)$ with the same total parity,
   it is in GUE with $ d_t=2 $ in the total parity sector.
   So we recovered the results listed in Table III for  even $N$ in this inter-color scheme.

 Note that if one do the exact diagonalization in the  $Q_{12}$ basis,
 then it mixes the two completely un-related parity sector $(Q_1,Q_2)$ and $(Q_1+1,Q_2+1)$,
 so one will find the ELS maybe something similar to the Poisson statistics shown in
 Fig.\ref{color1122hy}d.
 So only when doing exact diagonalization in a given parity sector  $ ( Q_1,Q_2 ) $,
 the ELS will show its real face: GOE.


In fact, there is a more straightforward way to do the classification and also find the degeneracy.
The parity of number operator $Q_a$, $a =1,2$ can be written as
$(-1)^{Q_a}=(i\chi_{a,1}\chi_{a,2})(i\chi_{a,3}\chi_{a,4})\cdots(i\chi_{a,N-1}\chi_{a,N})$.
Because the color 1 $ \chi_{1,i} $ is completely real
and color 2 $ \chi_{2,i} $ is completely imaginary,
we have $K(-1)^{Q_a}K=i^{N/2}(-1)^{Q_a}, a=1,2 $.
When $N\pmod 4=0$, 
the anti-unitary operator $K$ keeps both parities and $K^2=1$ which tell the ESL is GOE.
However, when $N\pmod 4=2$, 
$K$ maps $(Q_1,Q_2)$ to $ (Q_1+1,Q_2+1)$ with the same total parity.
So it is in GUE with $ d_t=2 $ in the total parity $ Q_{12} $ sector.

{\sl (b) N is odd }

When $N $ is odd, the exact diagonalization is most conveniently
and economically done without adding any Majorana fermions at $\infty$.
The classifications can be done either without adding or adding.
So the two theoretical approaches are complementary to each other
and should lead to the same answers when the degeneracy is counted carefully and correctly.

{\sl (b1) Classification in the minimum Hilbert space without adding Majorana fermions at $\infty$. }

Even in this case, there is no need to add extra decoupled Majorana fermions at $ \infty $.
Although $ Q_1, Q_2 $ makes no sense anymore, but $ Q_{12} $ is still well defined,
and one can do the exact diagonalization in the minimal Hilbert space
with just one conserved quantity $Q_{12}$.
Because $(-1)^{Q_{12}}=(i\chi_{1,1}\chi_{2,1})(i\chi_{1,2}\chi_{2,2})\cdots(i\chi_{1,N}\chi_{2,N})$
is always real, $K$ keeps the parity and $K^2=1$, which tell the ESL is GOE.
Since $ P_{12} Q_{12} P^{-1}_{12}= N-Q_t $,
$P_{12}$ maps $ Q_{12} $ to $ Q_{12}+1 $, so $ d_{t}=1+1 $.

{\sl (b2) Classification in the enlarged Hilbert space by adding Majorana fermions at $\infty$. }

In the following, we will do the classification and also find the degeneracy
in the enlarged Hilbert space by adding decoupled Majorana fermions at $\infty$.
In fact, as shown in (b1), this is not necessary in the inter-color scheme and in the balanced case.
However, it is constructive to do it here to compare with the intra-color scheme done in the main text.

As shown in Sec. III B, one can add $\chi_{1,N+1}=\chi_{1 \infty}$ and $\chi_{2,N+1}=\chi_{2\infty}$
to make the parity $\tilde{Q}_1$ conservation in the color 1
and $\tilde{Q}_1$ color 2 respectively and explicitly.
By adding the two Majorana fermions at $\infty$,
one doubles the Hilbert space, and also generates one more conserved parity.
In this inter-color scheme, it is convenient to take $(Q_{12},\tilde{Q}_{12})$ as a complete set,
where $Q_{12}$ is the total parity without adding the two Majorana fermions,
$\tilde{Q}_{12}=Q_{12}+n_{12 \infty}$ is the total parity including the two added Majorana fermions.

There are also two corresponding operators $P_{12}$, $P^2_{12}=(-1)^{N_c-1}$ and
$ \tilde{P}_{12}=P_{12} \chi_{1 \infty}$, $\tilde{P}^2_{12}=(-1)^{N_c} $,
where $ N_c=(N+1)/2$ as defined in  Sec. III B.
One can work out how the two operators act on the two conserved quantities:
$ P_{12} Q_{12} P^{-1}_{12}= N-Q_{12}$,
$P_{12} \tilde{Q}_{12} P^{-1}_{12}= N-\tilde{Q}_{12} + 2 n_{12 \infty } $
and $ \tilde{P}_{12} Q_{12} \tilde{P}^{-1}_{12}= N-Q_{12}$,
$\tilde{P}_{12} \tilde{Q}_{12} \tilde{P}^{-1}_{12}= N+1-\tilde{Q}_{12}  $.
Unfortunately, none of the two keeps the parity  $(Q_{12}, \tilde{Q}_{12})$.
One may try to use any combinations of $ P_{12}$, $\tilde{P}_{12} $ and $ R_{12}$, $\tilde{R}_{12} $
to construct relevant operators.
For example, one can try $ P= K P_{12} \tilde{P}_{12} = K \chi_{1 \infty} $,
then $ P Q_{12} P^{-1}= Q_{12}, P \tilde{Q}_{12} P^{-1}= 1-\tilde{Q}_{12} $,
so it still does not work.
However, if removing  $ \chi_{1 \infty} $ from $ P $,
then just $ K = P \chi_{1 \infty} $ along does the job.
Because $ K^2=1 $, so it is in GOE.

In fact, more straightforwardly, because both $(-1)^{Q_{12}}$ and $(-1)^{\tilde{Q}_{12}}$ are real, so
  $K$ keeps both parities and $K^2=1$ which tells the ESL is GOE.

  Note also that $\tilde{P}_{12}$ maps $(Q_{12}, \tilde{Q}_{12})$ to $(Q_{12}+1,\tilde{Q}_{12})$,
  so $d=1$ in  the minimal Hilbert space with just one conserved quantity $ Q_{12} $,
  but $ d_t=1+1 $ in the enlarged Hilbert space with a given parity of $\tilde{Q}_{12}$, so it
  can not be observed in the exact diagonalization done in the minimal Hilbert space  $ Q_{12} $.

  For $ N $ odd, using this inter-color $(Q_{12},\tilde{Q}_{12})$ scheme,
  we recover the Table III achieved in the main text
  in the intra-color  $(\tilde{Q}_1,\tilde{Q}_2)$ scheme.

\subsection*{B. 4-colored SYK models: 10-fold way and operator $P_{12}$ ($P_{34}$)
across the first two (the other two) colors and $P=K P_{12} P_{34}$.}

Just like the 2-colored SYK model discussed above,
   one can introduce $ N $ complex fermions from the first two colors
   $ c_i= ( \chi_{1i}-i \chi_{2i} )/\sqrt{2},
   c^{\dagger}_i= ( \chi_{1i}+i \chi_{2i} )/\sqrt{2} $
   and its number operator $  Q_c= \sum_i c^{\dagger}_i  c_i $.
   One define the anti-unitary particle-hole symmetry operator  to be
$P_{12}= K \prod^{N}_{i=1} ( c^{\dagger}_i + c_i )$.
In fact, $ R_{12}= K \prod^{N}_{i=1} ( c^{\dagger}_i -c_i ) $ work equally well,
but it will not lead to new symmetry.
It is easy to show $ P^2_{12}= (-1)^{\lfloor N/2 \rfloor} $.
One can also show that $  P_{12} c_i P_{12}= \eta c^{\dagger}_i$, $P_{12} c^{\dagger}_i P_{12}= \eta c_i$,
$P_{12} \chi_{ai} P_{12}= \eta \chi_{ai}$, $a=1,2$,
where $ \eta=(-1)^{\lfloor (N-1)/2 \rfloor} $.

One can also introduce $ N $ complex fermions from the other two flavors
$  d_i= ( \chi_{3i}-i \chi_{4i} )/\sqrt{2}$,
$d^{\dagger}_i= ( \chi_{3i}+i \chi_{4i} )/\sqrt{2} $ and
   its number operator $  Q_d= \sum_i d^{\dagger}_i  d_i $.
One can also define the similar anti-unitary operator $ P_{34} $ (or $ R_{34} $).
It is easy to see that $ P_{34} $ or $ R_{34} $ can do the same job,
but can not provide new information.
Of course, one can group differently such as $ P_{13}, P_{24} $ or $ P_{14}, P_{23} $,
and they should lead to the same answers.

It can be shown that $ P_{12} $ (also $ P_{34} $)
anti-commutes with the Hamiltonian $ \{ P_{12}, H_{1234} \}=0 $.
The total number of  fermions $ Q_t= \sum_i (c^{\dagger}_i  c_i + d^{\dagger}_i  d_i ) = Q_c + Q_d $
   is not a conserved quantity, but its parity $ (-1)^{Q_t} $ is.
   In fact, the parity $ (-1)^{Q_c} $ and $  (-1)^{Q_d} $  are separately conserved.
We have $ P_{12} Q_c P^{-1}_{12}= N-Q_c$, $P_{12} Q_d P^{-1}_{12}= Q_d $,
   so it maps to the same (opposite) parity in the $ Q_c $ sector when $ N $ is even (odd).
   Similarly,  $ P_{34} Q_c P^{-1}_{34}= Q_c$, $P_{34} Q_d P^{-1}_{34}= N- Q_d $.


     One can also define another anti-unitary operator as
\begin{equation}
    P=K \prod^{N}_{i=1} ( c^{\dagger}_i + c_i )  ( d^{\dagger}_i + d_i ) =P_{12} P_{34}  K
\label{p1234}
\end{equation}
   which can be contrasted to the similar operator in the 2-colored case Eq.\ref{p1122}.
   It is easy to show that $ P^2= (-1)^N $ and $ [ P, H_{1234} ]=0 $.
   It is also easy to see that $ P Q_c P^{-1}= N-Q_c$,  $P Q_d P^{-1}= N-Q_d $.
   Then $ P Q_t P^{-1}_{12}= N-Q_c + N- Q_d = 2N- Q_t $, so it always maps to the same total parity.

\begin{figure}[!htb]
\centering
\includegraphics[width=0.9\linewidth]{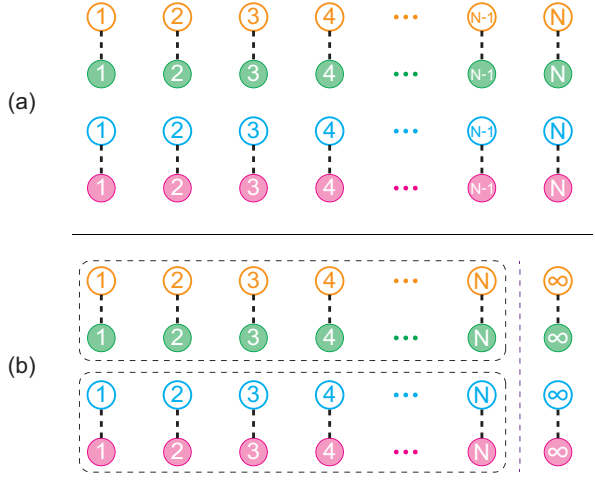}	
\caption{ The four colors with (a) $ N $ even and (b) $ N $ odd with the inter-color pairings across
 the color 1 and color 2, color 3 and color 4.
 In (b), the long vertical dashed line separate the system from the four Majorana fermions added at infinity.
 The Hilbert space is enlarged to $ 2 \times 2 =4 $ times.
 Simultaneously, there are also two more conserved parities.
 The two dashed boxes enclose the two conserved parities $ Q_{12} $ and $ Q_{34} $. }
\label{fourcolorapp.eps}
\end{figure}

{\sl (a) $ N $ is even }


  When combining $ P $ with $ P^2= (-1)^N $ and $ P_{12} $ with $ P^2_{12}= (-1)^{\lfloor N/2\rfloor} $,
  paying special attention to their action on the Hilbert space with a given parity in  $ (-1)^{Q_c} $ and $  (-1)^{Q_d} $,
  one can see that when $N \pmod 4=0, 2$, $ ( Q_c, Q_d ) $ maps to the same sector, so
  $ H_{1234} $ belongs to class BDI (chiral GOE ) or class CI (BdG)  respectively.
  The degeneracy $ d=1 $.
  These facts completely agree with the $N \pmod 4=0, 2$, case listed in Table IV.
  Note that one can also combine $ P $ in Eq.\ref{p1234} with any other $ P_{ij} $ or $ R_{ij} $ without affecting the results \cite{any}.
  Namely, one need to pick up just one representation $ ( P, P_{ij} ) $ or $ ( P, R_{ij} ) $.

  As alerted earlier, one can introduce $ N $ complex fermions from color 2 and 3,
  $  f_i= ( \chi_{3i}-i \chi_{2i} )/\sqrt{2}$,
  $ f^{\dagger}_i= ( \chi_{3i}+i \chi_{2i} )/\sqrt{2} $ and
  its number operator $ Q_{23}= Q_f= \sum_i f^{\dagger}_i  f_i $.
  One can also define the similar anti-unitary operator $ P_{23} $ (or $ R_{23} $).
  Of course,  $ Q_{23} $ enclosed in the box in Fig.\ref{fourcolorapp.eps}(a) is also a conserved parity.
  When $ N $ is even,  it also commutes with $  ( Q_{12}, Q_{34} ) $.
  Although it may not be conveniently  expressed in terms of the two groups of complex fermions $ c_i, d_i $,
  they can be conveniently expressed in terms of Majorana fermions of color 2 and 3.
  One can show that acting on it by $ P $ and $ P_{12} $ does not affect the results above
  achieved with $  ( Q_{12}, Q_{34} ) $ only.

  Now we find two anti-unitary operators, one commuting, another anti-commuting with $ H_{1234} $.
  From the two anti-unitary operators, one can define the chirality operator
  $ \Lambda= P_{12} P= P_{34}  K= \chi_{3,1} \chi_{3,2} \cdots \chi_{3, N}  $
  which is nothing but proportional to the parity operator $ i^{-N/2} (-1)^{Q_3} $ of the color 3.
  It is a unitary operator anti-commuting with the Hamiltonian $ \{ \Lambda, H_{1234} \}=0 $
  and also keeps all the parities $ (Q_{12}, Q_{23}, Q_{34} ) $.
  Of course, $ \Lambda= P_{12} K =\chi_{1,1} \chi_{1,2} \cdots \chi_{1, N} = i^{-N/2} (-1)^{Q_1}$
  works equally well as the unitary chirality operator.


  In fact, there is a more straightforward way to do the classification and also find the degeneracy.
  Because all the three parities $  ( Q_{12}, Q_{23}, Q_{34} ) $ are real, so $ T_{+}=K$ keep all and $K^2=1$.
  Because the Hamiltonian is real, so $ [K, H]=0 $.
  Note that we also have the chiral (mirror) symmetry $\{(-1)^{Q_a},H\}=0$ where $(-1)^{Q_a}=(i\chi_{a,1}\chi_{a,2})\cdots(i\chi_{a,N-1}\chi_{a,N})$
  and $a=1,2,3,4$. Thus we can construct the other anti-unitary operator as  $ T_{-}=K(-1)^{Q_a}$ which anti-commute
  with the Hamiltonian.
  From the fact $[(-1)^{Q_a},(-1)^{Q_b}]=0$, one can see that $ T_{-} $ keeps all the three parities.
  Since $(-1)^{Q_a}$ have the same sign as $i^{N/2}$, it is easy to see $ T^2_{-}=(-1)^{N/2}$.
  Combining $ T_{+}=K, T^2_{+}=1 $ and $ T_{-}=K(-1)^{Q_a}, T^2_{-}=(-1)^{N/2} $, we conclude that
  When $N \pmod 4=0$, it is class BDI; When $N \pmod 4=2$, it is class CI.

{\sl (b) $ N $ is odd }

Just similar to the 2-colored case, when $N$ is odd,
the exact diagonalization is most conveniently and economically done
without adding any Majorana fermions at $ \infty $.
The classifications can be done either without adding and adding,
so the two theoretical approaches are complementary to each other and should lead to the same answers.

{\sl (b1) Classification in the minimum Hilbert space without adding Majorana fermions at $ \infty $. }

Even in this case, there is no need to add extra Majorana fermions at $ \infty $.
There are still 3 conserved quantities $ Q_{12}, Q_{23},  Q_{34} $,
but  $ Q_{23} $ does not commute with $ ( Q_{12},  Q_{34} ) $ anymore,
so  we still do the exact diagonalization in
the minimal Hilbert space with just two conserved quantities $(Q_{12},Q_{34})$.
Because both $  Q_{12} $ and $ Q_{34}$ are real,
complex conjugate $ K $ keeps $ (Q_{12},Q_{34})  $ and $K^2=1$.
Again the Hamiltonian is real, so $ [K,H]=0 $.
These facts tell that the ESL is GOE. Note that $(-1)^{Q_a}$  is not defined for odd $ N $.
As said above, $ P $ maps $ ( Q_{12}, Q_{34} ) $ to $ (Q_{12}+1,  Q_{34}+1) $,
so level degeneracy is $ d=1 $ in the minimum Hilbert space with given $ ( Q_{12}, Q_{34} ) $,
and level degeneracy is $ d_{t}=2 $ in the total parity $ Q_t=  Q_{12} + Q_{34} $ sector.
This $ d_{t}=2 $ can be seen in the exact diagonalization in a given total parity $ Q_t $.
These results recover those listed in Table IV.

{\sl (b2) Classification in the enlarged Hilbert space by adding Majorana fermions at $ \infty $. }

In the following, we will do the classification and also find the degeneracy
in the enlarged Hilbert space by adding decoupled Majorana fermions at infinity.
In fact, as shown in (b1), this is not necessary
in the inter-color scheme and in the balanced case.
However, it is constructive to do it here to compare with that done in the main text.

As shown in Sec. V B,
one can add $ \chi_{a,N+1} =\chi_{a \infty}$, $a=1,2,3,4$,
to make the parity conservation in
$(\tilde{Q}_{12},\tilde{Q}_{23},\tilde{Q}_{34},Q_t)$ explicitly \cite{shift}.
Then one can repeat the procedures as in even $ N $ case in (a) with $ N \rightarrow N+1 $.
In this inter-color scheme, it is more convenient to take
$ (\tilde{Q}_{12},Q_{12},\tilde{Q}_{34},Q_{34}  ) $
as the complete set which is complementary to the set
$ (\tilde{Q}_{12},\tilde{Q}_{23},\tilde{Q}_{34},Q_t ) $ used in the main text \cite{shift}.
In this set, $ ( Q_{12}, Q_{34}  ) $  is the parity without adding the four Majorana fermions
   ( enclosed in the two boxes in Fig.\ref{fourcolorapp.eps}b )  and
   $ \tilde{Q}_{12}=Q_{12} + n_{12 \infty}$,  $\tilde{Q}_{34}=Q_{34} + n_{34 \infty}  $
   is the parity including the four Majorana fermions.
One can also construct two new operators
$ \tilde{P}_{12}=P_{12} \chi_{1 \infty}$, $\tilde{P}_{34}=P_{34} \chi_{3 \infty} $
   with $ \tilde{P}^2_{12}=\tilde{P}^2_{34}=(-1)^{[\lfloor(N+1)/2\rfloor} $.
   They lead to a new composite operator $ \tilde{P}= K \tilde{P}_{12}\tilde{P}_{34} $ with $ \tilde{P}^2=(-1)^{N+1} =1 $.

  Unfortunately, similar to the 2-colored case discussed above, none of the these operators keep the complete set of the parities
  $ (  Q_{12}, \tilde{Q}_{12}, Q_{34}, \tilde{Q}_{34} ) $ in the same sector.
  One can try to use $ P_{12}, \tilde{P}_{12}, R_{12}, \tilde{R}_{12} $ for color 1 and 2,
  $ P_{34}, \tilde{P}_{34}, R_{34}, \tilde{R}_{34} $ for color 3 and 4 to
  construct relevant operators in this inter-color scheme.
  Taking the experience from the 2-colored case, it turns out that just $ K $ along does the job.
  Because $ K^2=1 $, it is in GOE.

  In fact, more straightforwardly,
  since all the 4 parities of $ (  Q_{12}, \tilde{Q}_{12}, Q_{34}, \tilde{Q}_{34} ) $ are real,
  so $K$ keeps all the parities and $K^2=1$ which tells the ESL is GOE.
  Of course, all the 4 colors individual parities still anti-commute with the Hamiltonian $ \{ (-1)^{\tilde{Q}_a}, H \}=0, i=1,2,3,4 $,
  but none can keep all the cross parities. For example,
  $(-1)^{\tilde{Q}_1}$ keeps $\tilde{Q}_{12}$, but changes $Q_{12}$.
  Note that $ (-1)^{Q_a} $ is not chiral operator for odd $ N $.

  Note also that $ \tilde{P} $ maps $  ( Q_{12}, Q_{34}, \tilde{Q}_{12}, \tilde{Q}_{34}  ) $
  to $  ( Q_{12}+1, Q_{34}+1, \tilde{Q}_{12}, \tilde{Q}_{34}  ) $
  which has the same total parity $ Q_t= Q_{12} + Q_{34} $,
  so $ d=1 $ at a given parity  $  ( Q_{12}, Q_{34}  ) $ and $ d_t=2 $ in the total parity $ Q_t $,
  consistent with that listed for  $ N \pmod 4=1, 3 $ in Table IV.
  Obviously, this double degeneracy can be observed in the exact diagonalization done in the total parity $ Q_t $ sector shown in
  Fig.\ref{color1234hybrid} (a) and (c).
  Of course, this scheme can not be even used in the imbalanced case.


 Finally, we conclude that the biggest advantage to use the inter-color scheme is that
 the Hamiltonian is made real, if all the  conserved quantities are real,
 then the bulk ESL must be GOE. This is why 7 out of 8 cases in Table III and Table IV are real.
 The only exception is that in the two color case with $N \pmod 4=2$, as shown in Appendix A-a,
 both $ Q_1 $ and $ Q_2 $ are imaginary, so it is in GUE.
 As shown in \cite{KAM,u1largen,gold,comment,strongED} on Dicke model,
 $ K $ is the only relevant anti-unitary operator
 which commutes the Dicke Hamiltonian, so $ K^2=1 $ only leads to GOE for the Dicke model.

 Unfortunately, as said before, this inter-color scheme can not be extended to the imbalanced case.
 When generalizing the method used in the main text to all the possible imbalanced cases with $ q=4 $,
 we find all the 10 classes in the 10-fold way classifications \cite{kamsykcolorim}.


\subsection*{C. Classifications of the hybrid 2- and 4- colored SYK  models in the inter-color scheme}

 It turns out that the inter-color scheme may be used to reach
 the classifications of the hybrid 2- and 4- colored SYK  models
 quicker than the intra-color scheme used in the main text.
 The complex conjugate operator $K$ is the operator who does most of the job.

{\sl (a) hybrid 2-colored case.}

For the representation used in Fig.\ref{twocolorapp.eps},
the hybrid Hamiltonian $H^{Hb}_{1122}$ in Eq.\ref{gr2hyv} is real
and the only conserved quantity is total parity $(-1)^{Q_t}$.
No matter $N$ is even or odd, $Q_t= Q_{12} $ is always real and thus $(-1)^{Q_t}$ is also real.
Then $[K, H^{Hb}_{1122}]=0$ and $[K,(-1)^{Q_t}]=0$.
Of course, $K^2=1$, so its ESL is  GOE at any ratio $ J/K $.

{\sl (b) hybrid 4-colored case.}

Since the quadratic term $ i\sum_{i,j;a<b} K^{ab}_{ij} \chi_{ai} \chi_{bj} $
contains all kinds of inter-color couplings,
The inter-color scheme shown in Fig.\ref{fourcolorapp.eps} can not be used to make it real.
In fact, it has no symmetry, so the ESL of the total Hamiltonian $ H^{Hb}_{1234}$ in Eq.\ref{gr4all} is in GUE at $ J/K\sim 1$.
If one changes the quadratic term to
$i\sum K_{ij}(\chi_{1i}\chi_{2j}+\chi_{3i}\chi_{4j})$,
then just like the hybrid 2-color case, it also becomes real,
the modified hybrid 4-colored SYK model must
also be in GOE at $ J/K\sim 1$.


\begin{thebibliography}{999}
\bibitem{uncolor}
The history of the colored SYK versus SYK is just opposite to that of the colored Tensor versus uncolored Tensor model.
  Gross and Rosenhaus \cite{Gross} put colors on the orginal SYK model. While Klebanov and Tarnopolsky \cite{tensor3}
  un-colored the original  colored tensor (Gurau-Witten) model.
\bibitem{also}
The simplest case is the $N \pmod 8=0$ case in the SYK,
$ P^2 =1 $, it also maps $ Q $ to the same sector,
so it is in GOE with $ d=1 $ at a given parity $ Q $.
However, there is another parity sector, the two parity sectors are completely un-correlated,
so it is not known which parity sector contains the ground state.
However, if doing exact diagonalization in both parity sectors, one get something similar to Poisson.
The rigrous RMT of this mixed 2 un-correlated random matrix emsembles was presented in Sec.II-B.
See also Appendix A.
\bibitem{comp2}
This is similar to the $N \pmod 8=6$ case in the SYK,
$ P^2 =-1 $, it also maps $ Q $ to the opposite sector $ Q+1 $,
so it is in GUE with $ d=1 $ at a given parity $ Q $.
However, it has the $ d_t=2 $ double degeneracy when considering both parity sectors.
See also Appendix A.
\bibitem{any}
To some extent, this is similar to the case central potential quantum mechanical problem,
one can choose the complete set $ J^2, J_z $ or  $ J^2, J_x $ or $ J^2, J_y $.
but $ J_i, i=x,y,z $ do not commute with each other.
\bibitem{pmpz}
Of course, in the $ q=4 $ case,
one can equally use $ P_m $ operator (plus $ P_z $) to characterize the symmetry of $ H_4 $
reach the same conclusion as that using $P$ (plus $ P_z $).
\bibitem{shift}
Note that due to adding Majorana fermions at $ \infty $,
there is a shift of notation here from the main text to the two appendics when $ N $ is odd.
In the two color case, $(Q_1,Q_2)$ in the main text to $ ( \tilde{Q}_1, \tilde{Q}_2 ) $
in the appendix A.
In the 4 color case, $ ( Q_{12}, Q_{23}, Q_{34}, Q_{0t} ) $  used in main text to
$ ( \tilde{Q}_{12}, \tilde{Q}_{23}, \tilde{Q}_{34}, Q_t ) $ used in the appendix B.
\end{thebibliography}
\end{document}